\documentclass[prd,nofootinbib,showpacs,preprint]{revtex4-1}
\usepackage[T1]{fontenc}
\usepackage{amsmath,amssymb}
\usepackage{epsfig}
\usepackage{graphicx}
\usepackage[usenames,dvipsnames]{color}
\usepackage{slashed}
\usepackage[colorlinks,citecolor=blue]{hyperref}
\usepackage{pdfpages}
\usepackage{color}
\begin{document}
%\preprint{IP/BBSR/2015-4}
\title{Non-zero $\theta_{13}$ with Unbroken $\mu-\tau$ Symmetry of the Active Neutrino Mass Matrix in the Presence of a Light Sterile Neutrino}

\author{Debasish Borah}
\email{dborah@iitg.ernet.in}
\affiliation{Department of Physics, Indian Institute of Technology Guwahati, Assam 781039, India}
\begin{abstract}
We revisit the possibility of generating non-zero reactor mixing angle in a scenario where there is a sterile neutrino at the eV scale apart from the usual three sub-eV scale active neutrinos. We show that the $3\times3$ active neutrino mass matrix can possess a $\mu-\tau$ symmetry and can still be consistent with non-zero value of the reactor mixing angle $\theta_{13}$, if this $\mu-\tau$ symmetry is broken in the sterile neutrino sector. We first propose a simple model based on the discrete flavour symmetry $A_4 \times Z_3 \times Z^{\prime}_3$ to realise such a scenario and then numerically evaluate the complete $3+1$ neutrino parameter space that allows such a possibility. We show that, such a possibility of generating non-zero $\theta_{13}$ can in general, remain valid even if the present $3+1$ neutrino global fit data get ruled out by future experiments. We also discuss the possible implications at neutrinoless double beta decay $(0\nu \beta \beta)$ experiments in view of the latest results from KamLAND-Zen experiment.
\end{abstract}
\maketitle

\section{Introduction}
Origin of non-zero neutrino masses and large leptonic mixing \cite{PDG, kamland08, T2K, chooz, daya, reno, minos} has been one of the longstanding puzzles in particle physics. Although the 2012 discovery of the Higgs boson at the Large Hadron Collider (LHC) has confirmed the validity of the Standard Model (SM) of particle physics, the model however fails to account for the observations in the neutrino sector. This has given rise to several beyond standard model (BSM) physics proposals that can generate non-zero neutrino masses and mixing, in agreement with experimental observations. The $3\sigma$ global fit values of neutrino oscillation parameters that have appeared in the recent analysis of \cite{schwetz14}  and \cite{valle14}  are shown in table \ref{tab:data1}.
\begin{center}
\begin{table}[htb]
\begin{tabular}{|c|c|c|c|c|}
\hline
Parameters & NH \cite{schwetz14} & IH \cite{schwetz14} & NH \cite{valle14} & IH \cite{valle14} \\
\hline
$ \frac{\Delta m_{21}^2}{10^{-5} \text{eV}^2}$ & $7.02-8.09$ & $7.02-8.09 $ & $7.11-8.18$ & $7.11-8.18 $ \\
$ \frac{|\Delta m_{31}^2|}{10^{-3} \text{eV}^2}$ & $2.317-2.607$ & $2.307-2.590 $ & $2.30-2.65$ & $2.20-2.54 $ \\
$ \sin^2\theta_{12} $ &  $0.270-0.344 $ & $0.270-0.344 $ &  $0.278-0.375 $ & $0.278-0.375 $ \\
$ \sin^2\theta_{23} $ & $0.382-0.643$ &  $0.389-0.644 $ & $0.393-0.643$ &  $0.403-0.640 $ \\
$\sin^2\theta_{13} $ & $0.0186-0.0250$ & $0.0188-0.0251 $ & $0.0190-0.0262$ & $0.0193-0.0265 $ \\
$ \delta $ & $0-2\pi$ & $0-2\pi$ & $0-2\pi$ & $0-2\pi$ \\
\hline
\end{tabular}
\caption{Global fit $3\sigma$ values of neutrino oscillation parameters \cite{schwetz14, valle14}.}
\label{tab:data1}
\end{table}
\end{center}
Since only two mass squared differences are measured experimentally, the lightest neutrino mass is still unknown. Also the mass ordering is not settled yet, allowing both normal hierarchy (NH) as well as inverted hierarchy (IH). Cosmology experiments can however, put an upper bound on the lightest neutrino mass from the measurement of the sum of absolute neutrino masses $\sum_i \lvert m_i \rvert < 0.17$ eV \cite{Planck15}. Although the solar and atmospheric mixing angles $(\theta_{12}, \theta_{23})$ were known to have large values, the discovery of non-zero $\theta_{13}$ is somewhat recent \cite{T2K, chooz, daya, reno, minos}. The leptonic Dirac CP phase $\delta$ is not yet measured experimentally \footnote{A recent measurement hinted at $\delta \approx -\pi/2$ \cite{diracphase}.} though the global fit data indicate the best fit value as: $306^o$ (NH), $254^o$ (IH) \cite{schwetz14} and $254^o$ (NH), $266^o$ (IH) \cite{valle14}. If neutrinos are Majorana fermions, then two other CP phases appear, which do not affect neutrino oscillation probabilities and hence remain undetermined in such experiments. They can however be probed at experiments looking for neutrinoless double beta decay $(0\nu \beta \beta)$.

Neutrinos remain massless in the SM due to the absence of the right handed neutrino which is required in order to allow Yukawa couplings between the neutrinos and the Higgs field. Even if the right handed neutrinos are introduced by hand, one needs to fine tune the dimensionless Yukawa couplings to the level of $10^{-12}$ in order to allow sub eV neutrino masses. One can generate a tiny Majorana neutrino mass through dimension five Weinberg operators in the SM \cite{weinberg} in an effective theory framework. Several BSM proposals for the realisation of such an operator within a renormalisable theory have appeared in the literature in the last few decades which are more popularly known as seesaw mechanisms. Apart from the dynamical origin of tiny neutrino masses, the observed pattern of leptonic mixing has also been a puzzle particularly due to the large mixing angles. This is in sharp contrast with the quark sector where the mixing angles are very small. Prior to the discovery of non-zero $\theta_{13}$, such large leptonic mixing angles (solar and atmospheric) were consistent with a class of neutrino mass matrices obeying $\mu-\tau$ symmetry \footnote{For a recent review, please see \cite{xing2015}.}. This class of models predicts $\theta_{13} = 0, \theta_{23} = \frac{\pi}{4}$ whereas the value of $\theta_{12}$ depends upon the particular model. Out of different neutrino mixing patterns that can originate from such a $\mu-\tau$ symmetric neutrino mass matrix, the Tri-Bimaximal (TBM) \cite{Harrison} mixing pattern received more attention in the neutrino model building studies. The TBM mixing predicts $\theta_{12}=35.3^o$. Such a mixing can be easily accommodated within popular discrete flavour symmetry models \cite{discreteRev}. Among them, the discrete group $A_4$ which is the group of even permutations of four objects, can reproduce the TBM mixing in the most economical way \cite{A4TBM, A4TBM1}. Since the latest neutrino oscillation data is not consistent with $\theta_{13}=0$ and hence TBM mixing, one has to go beyond the minimal $\mu-\tau$ symmetric framework. Since the measured value of $\theta_{13}$ is small compared to the other two, one can still consider the validity of $\mu-\tau$ symmetry at the leading order and generate non-zero $\theta_{13}$ by adding small $\mu-\tau$ symmetry breaking perturbations. Such corrections can originate from the charged lepton sector or the neutrino sector itself like for example, in the form of a new contribution to the neutrino mass matrix. This has led to several works including \cite{nzt13, nzt13A4, nzt13GA,db-t2, dbijmpa, dbmkdsp, dbrk} within different BSM frameworks.

Another interesting but much less explored idea to generate non-zero $\theta_{13}$ is by allowing the mixing of three active neutrinos with a eV scale sterile neutrino \cite{sterilemutau0, sterilemutau, sterilemutau1, sterilemutau2}. For a review of light sterile neutrinos at eV scale, please refer to \cite{whitepaper}. Such light sterile neutrinos received lots of attention after the LSND accelerator experiment reported anomalies in the measurement of antineutrino flux \cite{LSND1} which was later supported by results from the MiniBooNE experiment \cite{miniboone}. Reactor neutrino experiments \cite{react} as well as gallium solar neutrino experiments \cite{gall1,gall2} also discovered similar anomalies. These anomalies require the presence of a light sterile neutrino at eV scale with non-trivial mixing with the active neutrinos as presented in the global fit studies \cite{globalfit, globalfit2}. Although cosmology experiments like Planck \cite{Planck15} leave no room to accommodate one additional light sterile neutrino within the standard $\Lambda$CDM model of cosmology, one can evade these tight bounds by considering the presence of some new physics. For example, additional gauge interactions in order to suppress the production of sterile neutrinos through flavour oscillations were studied recently by the authors of \cite{kopp}. Recently, the IceCube experiment at the south pole has excluded the the $3+1$ neutrino parameter space mentioned in global fit data \cite{globalfit} at approximately $99\%$ confidence level \cite{icecube1}. However, in the presence of non-standard interactions, the $3+1$ neutrino global fit data can remain consistent with the IceCube observations \cite{marfatia}. Therefore, there is still room for the existence of an eV scale sterile neutrino within some specific BSM frameworks that can provide a consistent interpretation of experimental data. Here we intend to study the consequence of such $3+1$ neutrino scenario on the $\mu-\tau$ symmetry in a way first discussed by the authors of \cite{sterilemutau, sterilemutau1, sterilemutau2}. We in fact point out that, such a scenario of breaking $\mu-\tau$ symmetry from the sterile neutrino sector can remain valid even if the present and future neutrino experiments conclusively rule out the $3+1$ global fit data \cite{globalfit}. We also propose a model to realise such a scenario based on $A_4 \times Z_3 \times Z^{\prime}_3$ flavour symmetry.

In such a framework, the light neutrino mass matrix is $4\times4$ and non-zero $\theta_{13}$ is possible even if the $3\times3$ active neutrino block preserves a $\mu-\tau$ symmetry, whereas the sterile neutrino sector breaks it. This was first proposed by the authors of \cite{sterilemutau0} and was discussed in more details in \cite{sterilemutau, sterilemutau1, sterilemutau2} later on. This can have very interesting implications for neutrino model building in the presence of flavour symmetries. Although simple analytical understanding of such a framework have been presented in one of the recent works \cite{sterilemutau2}, a complete numerical analysis is still missing. To be more specific, if we demand the active neutrino block of the $4\times4$ light neutrino mass matrix to possess an underlying $\mu-\tau$ symmetry, it can have interesting implications for the neutrino parameters. In the minimal $A_4$ realisation of such $\mu-\tau$ symmetric or TBM type active neutrino mass matrix, one has even more restrictions on the elements of the active neutrino mass matrix. This can restrict the active-sterile mixing as well as the CP phases to some specific values that can undergo further scrutiny at ongoing oscillation experiments \cite{sterileExpt16}. We also study the implications of these scenarios at $0\nu\beta \beta $ experiments.

This paper is organised as follows. In section \ref{sec1}, we discuss the basics of $\mu-\tau$ symmetry and its implications in $3+1$ neutrino framework. In section \ref{sec2}, we discuss a $A_4 \times Z_3 \times Z^{\prime}_3$ realisation of the $4\times 4$ light neutrino mass matrix that preserves a $\mu-\tau$ symmetry in the $3\times3$ active block and also discuss the issue of vacuum alignment and other interesting phenomenology of the flavon fields. We discuss the procedures followed in numerical calculations in section \ref{sec3} and finally summarise our results and conclusion in section \ref{sec4}.

\section{$\mu-\tau$ Symmetry in $3+1$ Framework}
\label{sec1}
The $\mu-\tau$ symmetric mass matrices are symmetric under the interchange of $\mu \leftrightarrow \tau$. In the usual three neutrino scenario, the $3\times3$ mass matrix with $\mu-\tau$ symmetry can be written as
\begin{equation}
 M^{3\times3}_{\mu-\tau}=\left(\begin{array}{ccc}
 A&B&B\\
 B&C&D\\
 B&D&C
 \end{array}\right)
 \label{mutaugeneral}
\end{equation}
which is clearly symmetric with respect to the $2 \leftrightarrow 3$ or $\mu \leftrightarrow \tau$ interchange. Here, the neutrinos are assumed to be Majorana fermions having a complex symmetric mass matrix. The Pontecorvo-Maki-Nakagawa-Sakata (PMNS) leptonic mixing matrix is related to the diagonalising
matrices of neutrino and charged lepton mass matrices $U_{\nu}, U_{\ell}$ respectively, as
\begin{equation}
U_{\text{PMNS}} = U^{\dagger}_{\ell} U_{\nu}
\label{pmns0}
\end{equation}
Assuming the charged lepton mass matrix to be diagonal or equivalently $U_{\ell} = I$, one can find the leptonic mixing matrix just by diagonalising the above mass matrix \eqref{mutaugeneral}. It is straightforward to diagonalise the $\mu-\tau$ symmetric neutrino mass matrix to find two of the mixing angles as $\theta_{23} = \pi/4, \theta_{13} = 0$. The numerical value of the other mixing angle $\theta_{12}$ depends upon the relation between the parameters $A, B, C, D$ of the mass matrix. Similarly, one can also write down a $\mu-\tau$ symmetric mass matrix in the $3+1$ neutrino scenario
\begin{equation}
 M=\left(\begin{array}{cccc}
 A&B&B & F \\
 B&C&D & G\\
 B&D&C& G \\
 F & G & G& H \\
 \end{array}\right)
\end{equation}
which results in $\theta_{13}=0$ as long as $M_{\mu s} = M_{\tau s}=G$ is maintained. As shown in \cite{sterilemutau1, sterilemutau2}, one can generate non-zero value for the reactor mixing angle $\theta_{13}$ by introducing a breaking of $\mu-\tau$ symmetry in the sterile sector that is, $M_{\mu s} \neq M_{\tau s}$ while keeping the $3\times3$ active neutrino block $\mu-\tau$ symmetric. The authors of \cite{sterilemutau2} derived approximate analytical expressions for the active neutrino mixing angles as a function of sterile neutrino parameters, by considering the effective $3\times3$ neutrino mass matrix after the decoupling of the sterile neutrino. More specifically, the reactor mixing angle was derived as a function of $\mu-\tau$ symmetry breaking parameter $\Delta M = M_{\tau s} - M_{\mu s}$. Instead of deriving the approximate analytical formulas, here we investigate the constraints on $3+1$ neutrino parameters by imposing a $\mu-\tau$ symmetry in the $3\times3$ block. We also evaluate the deviation $\Delta M$ required to generate all the neutrino parameters within experimentally allowed range.

\section{$A_4 \times Z_3 \times Z^{\prime}_3$ Model for $3+1$ Neutrino Framework}
\label{sec2}
Several BSM frameworks have been proposed in order to generate three active and one sterile neutrino masses simultaneously near the eV scale \cite{sterilemutau1, sterilemutau2, sterileearlier1, sterileearlier3, sterileearlier5, sterileearlier6, sterileearlier7, sterileearlier8, sterileearlier9, sterileearlier10}. Recently, another model was proposed \cite{db164zero} which generates the $3\times3$ block of the $4\times 4$ light neutrino mass matrix through type II seesaw mechanism \cite{tii} whereas the active-sterile and sterile-sterile terms are generated by higher dimensional operators . Usually, there are two aspects of such model building efforts: (i) to find a dynamical origin of three active and one sterile neutrino masses around the eV scale along with non-trivial active-sterile mixing, (ii) to find a dynamical origin of the specific mixing patterns of the active-active and active-sterile sector. Here we mainly focus on the latter aspect and consider the origin of active and sterile mass scale from effective higher dimensional terms suppressed by a cut-off scale $\Lambda$. To be more specific, we consider a flavour symmetric model based on the discrete non-abelian group $A_4$ augmented by $Z_3 \times Z^{\prime}_3$ which predicts the specific structure of the $4\times 4$ light neutrino mass matrix in a natural and minimal way. It is also possible to propose a renormalisable version of this model by implementing a specific seesaw mechanism behind the origin of tiny neutrino masses. However, in this work we stick to the minimal field content required to generate the desired structure of lepton mass matrices and to guarantee the desired vacuum alignment. Therefore, we do not specify any particular seesaw mechanism and confine ourselves to discussing lepton masses through non-renormalisable terms in the superpotential.

The discrete group $A_4$ is the group of even permutations of four objects or the symmetry group of a tetrahedron. It has twelve elements and four irreducible representations with dimensions $n_i$ such that $\sum_i n_i^2=12$. These four representations are denoted by $\bf{1}, \bf{1'}, \bf{1''}$ and $\bf{3}$ respectively. The product rules for these representations are given in appendix \ref{appen2}. Here we consider a simple extension of the Altarelli-Feruglio model \cite{A4TBM1} in order to take the light sterile neutrino into account. The minimum field content required to arrive at the desired structure of the $4\times4$ light neutrino mass matrix is shown in table \ref{table1}. The transformations of the lepton doublets $l$, charged lepton singlets $e_R, \mu_R, \tau_R$ and Higgs doublets $H_{u,d}$ under the gauge symmetry are same as in the minimal supersymmetric standard model (MSSM). All other fields are singlet under the MSSM gauge symmetry. The flavon fields $\phi_E, \phi_{\nu}, \eta$ are enough to generate the $\mu-\tau$ symmetric active neutrino mass matrix along with a diagonal charged lepton mass matrix. The other flavon fields $\phi_S, \chi, \psi, \zeta$ are introduced in order to generate the $\mu-\tau$ symmetry breaking sterile neutrino sector.

\begin{table}[h]
\begin{tabular}{|c|c|c|c|c|c|c|c|c|c|c|c|c|c|c|}
\hline
& $l$ & $e_R$  & $\mu_R$ & $\tau_R$ & $\nu_s$ &$H_{u,d}$  &$\phi_E$  & $ \phi_{\nu}$  & $\eta, \bar{\eta}$  & $\phi_S$ & $\chi, \bar{\chi}$  & $\psi, \bar{\psi} $  & $\zeta, \bar{\zeta}$ &$\xi, \bar{\xi}$ \\ \hline
 $A_4$& 3 & $1$ & $1'$ & $1''$ & $1$  &$1$  & 3 &  3& 1 & 3 & $1$ & $1'$ & $1''$   & $1$\\
 $Z_3$& $\omega$ & $\omega^2$ & $\omega^2$ & $\omega^2$ & $1$  &$1$  & 1 & $\omega$ &  $\omega$  & $\omega$ & $\omega$ & $\omega$ & $\omega$ & 1 \\
 $Z^{\prime}_3$& $0$ & $0$ & $0$ & $0$ & $\omega$  &$0$  & 0 & $0$ &  0 & $\omega$ & $\omega$ & $ \omega$ & $ \omega$ & $\omega$ \\\hline

\end{tabular}
\caption{Transformation of the fields under $A_4 \times Z_3 \times Z^{\prime}_3$ symmetry of the model.}
\label{table1}
\end{table}

For the field content in table \ref{table1}, one can write down the corresponding superpotential similar to the way it was done in the original Altarelli-Feruglio model \cite{A4TBM1}. Here we write down the corresponding superpotential and derive the light neutrino mass matrix. It is straightforward to write down the superpotential as
\begin{align}
W  & \supset Y_e e_R l H_d \frac{\phi_E}{\Lambda}+Y_{\mu} \mu_R l H_d \frac{\phi_E}{\Lambda} +Y_{\tau} \tau_R l H_d \frac{\phi_E}{\Lambda} + (x_a \eta + \overline{x_a} \bar{\eta}) l H_u l H_u \frac{1}{\Lambda^2}  \nonumber \\
& + x_b l H_u l H_u \frac{\phi_{\nu}}{\Lambda^2}+ (x_c \chi + \overline{x_c} \bar{\chi}) l H_u \nu_s \frac{\phi_S}{\Lambda^2}+(x_d \psi + \overline{x_d} \bar{\psi}) l H_u \nu_s \frac{\phi_S}{\Lambda^2} \nonumber \\
& +(x_e \zeta + \overline{x_e} \bar{\zeta}) l H_u \nu_s \frac{\phi_S}{\Lambda^2} + (x_s \xi + \overline{x_s} \bar{\xi}) \nu_s \nu_s H_u H_d \frac{1}{\Lambda^2}+ \text{h.c.}
\end{align}
The above superpotential, apart from being invariant under the MSSM gauge symmetry as well as $A_4 \times Z_3 \times Z^{\prime}_3$ discrete symmetry also has a continuous symmetry $U(1)_R$ that contains the usual R-parity as a subgroup. In comparison to the model \cite{A4TBM1}, here we have an additional $Z^{\prime}_3$ symmetry in order to generate the desired active-sterile and sterile-sterile entries, without affecting the structure of the active neutrino mass matrix. This additional $Z^{\prime}_3$ symmetry prevents a tree level bare mass term $m_s \nu_s \nu_s$ for the sterile neutrino, which is otherwise allowed by the gauge symmetry of the model. Preventing this bare mass term is necessary because an eV scale bare mass term in the superpotential is unnatural. One can however consider an approximate $U(1)$ global symmetry which is only broken by the bare mass term. In such a case, $m_s$ can be naturally small (of eV scale, say) as $m_s \rightarrow 0$ helps in recovering the full $U(1)_S$ global symmetry of the superpotential. One can also extend it to $U(1)_S$ gauge symmetry by introducing additional fields required for anomaly cancelation, which we do not pursue here and stick to this minimal structure of the model. As shown in \cite{A4TBM1}, the vacuum alignments $\langle \phi_E \rangle =(v_E, 0, 0), \langle \phi_{\nu} \rangle =(v_{\nu}, v_{\nu}, v_{\nu}), \langle \eta \rangle = u$ give rise to a diagonal charged lepton mass matrix and a $\mu-\tau$ symmetric $3\times3$ active neutrino mass matrix. This $\mu-\tau$ symmetric $3\times3$ has a structure
\begin{equation}
 M^{3\times3}_{\mu-\tau}=\left(\begin{array}{ccc}
 a+2b/3&-b/3&-b/3\\
 -b/3&2b/3&a-b/3\\
 -b/3&a-b/3&2b/3
 \end{array}\right)
\label{mutauA4}
\end{equation}
where $ a = 2 x_a u \frac{v^2_u}{\Lambda^2}, b = 2 x_b v_{\nu} \frac{v^2_u}{\Lambda^2}$. The vacuum expectation value (vev) of $H_u$ is denoted by $v_u$. This form of the mass matrix can be easily derived by using the $A_4$ product rules given in appendix \ref{appen2} with the above choice of vacuum alignments. This particular $\mu-\tau$ symmetric active neutrino mass matrix gives rise to the TBM mixing pattern discussed before. Now, due to the presence of additional terms in the superpotential involving the sterile neutrino $\nu_s$, the light neutrino mass matrix is $4\times4$ having the following structure
\begin{equation}
 M=\left(\begin{array}{cc}
 M^{3\times3}_{\mu-\tau} &\vec{\alpha}\\
 \vec{\alpha}^T & m_s
 \end{array}\right)
\label{mutau44}
\end{equation}
where $\vec{\alpha}^T = (M_{es}, M_{\mu s}, M_{\tau s} )$ contains the active-sterile mixing elements. Assuming the simple vacuum alignments $\langle \phi_{S} \rangle =(v_{S}, 0, 0), \langle \chi, \psi, \zeta \rangle = u_{1,2,3}$ one can derive these mixing elements as
$$ M_{es} = x_c v_u \frac{u_1 v_S}{\Lambda^2}, \; M_{\mu s} = x_e v_u \frac{u_3 v_S}{\Lambda^2}, \; M_{\tau s} = x_d v_u \frac{u_2 v_S}{\Lambda^2} $$
Thus, the sterile neutrino sector can break the $\mu-\tau$ symmetry if $x_e u_3 \neq x_d u_2$, which is easy to achieve by different choices of dimensionless couplings $x_{d,e}$ and singlet vev's $u_{2,3}$. One can also achieve a $\mu-\tau$ symmetry breaking $\vec{\alpha}$ without introducing the singlet flavons $\psi, \zeta$ if the triplet $\phi_S$ has a vacuum alignment $\langle \phi_S \rangle = (v_{eS}, v_{\mu S}, v_{\tau S})$ with $v_{\mu S} \neq v_{\tau S}$.

\subsection{Vacuum Alignment}
The choice of vacuum alignment of the flavon fields required to achieve the desired structures of lepton mass matrices mentioned above can be realised only when additional driving fields are incorporated as discussed in \cite{A4TBM1}. Since those fields do not affect the general structure of the mass matrices, we have not incorporated them in the discussion above. The non-trivial vacuum alignment of the $\phi_E, \phi_{\nu}$ fields required to produce the specific structure of the charged lepton mass matrix and the $3\times 3$ block of the light neutrino mass matrix is realised by introducing three additional driving fields, as shown in \cite{A4TBM1}. Denoting these driving fields as $\phi^E_0, \phi^{\nu}_0, \eta_0$ which have similar $A_4 \times Z_3 \times Z^{\prime}_3$ transformations as $\phi_E, \phi_{\nu}, \eta$, we can write down the superpotential involving these driving fields as
\begin{align}
W_d &  \supset M_1 (\phi^E_0 \phi_E) + \alpha_1 (\phi^E_0 \phi_E \phi_E) + \alpha_2 (\phi^{\nu}_0 \phi_{\nu} \phi_{\nu}) + \alpha_3 \bar{\eta} (\phi^{\nu}_0 \phi_{\nu}) \nonumber \\
& + \alpha_4 \eta_0 (\phi_{\nu} \phi_{\nu}) + \alpha_5 \eta_0 \eta^2 + \alpha_6 \eta_0 \eta \bar{\eta} + \alpha_7 \eta_0 \bar{\eta}^2
\end{align}
In the vacuum alignment, the presence of $\bar{\eta}$ plays a very non-trivial role, as discussed in \cite{A4TBM1}. In the above superpotential, this field $\bar{\eta}$ is considered to be the combination of $\eta, \bar{\eta}$ that couples to $(\phi^{\nu}_0 \phi_{\nu})$. Minimisation of the scalar potential in the supersymmetric limit gives rise to the following conditions from the driving sector
\begin{equation}
\frac{\partial W_d}{\partial \phi^{E}_{01}} = M_1 \phi_{E1} +\frac{2}{3} \alpha_1 (\phi^2_{E1}-\phi_{E2} \phi_{E3}) =0
\end{equation}
\begin{equation}
\frac{\partial W_d}{\partial \phi^{E}_{02}} = M_1 \phi_{E3} +\frac{2}{3} \alpha_1 (\phi^2_{E2}-\phi_{E1} \phi_{E3}) =0
\end{equation}
\begin{equation}
\frac{\partial W_d}{\partial \phi^{E}_{03}} = M_1 \phi_{E2} +\frac{2}{3} \alpha_1 (\phi^2_{E3}-\phi_{E1} \phi_{E2}) =0
\end{equation}
\begin{equation}
\frac{\partial W_d}{\partial \phi^{\nu}_{01}} = \alpha_3 \bar{\eta} \phi_{\nu 1} +\frac{2}{3} \alpha_2 (\phi^2_{\nu 1}-\phi_{\nu 2} \phi_{\nu 3}) =0
\end{equation}
\begin{equation}
\frac{\partial W_d}{\partial \phi^{\nu}_{02}} = \alpha_3 \bar{\eta} \phi_{\nu 3} +\frac{2}{3} \alpha_2 (\phi^2_{\nu 2}-\phi_{\nu 1} \phi_{\nu 3}) =0
\end{equation}
\begin{equation}
\frac{\partial W_d}{\partial \phi^{\nu}_{03}} = \alpha_3 \bar{\eta} \phi_{\nu 2} +\frac{2}{3} \alpha_2 (\phi^2_{\nu 3}-\phi_{\nu 1} \phi_{\nu 2}) =0
\end{equation}
\begin{equation}
\frac{\partial W_d}{\partial \eta_{0}} = \alpha_5 \eta^2 +\alpha_6 \eta \bar{\eta} + \alpha_7 \bar{\eta}^2+\alpha_4 (\phi^2_{\nu 1}+2\phi_{\nu 2} \phi_{\nu 3}) =0
\end{equation}
It is clear from the above minimisation conditions that the desired vacuum alignment $\langle \phi_E \rangle =(v_E, 0, 0), \langle \phi_{\nu} \rangle =(v_{\nu}, v_{\nu}, v_{\nu}), \langle \eta \rangle = u, \langle \bar{\eta} \rangle = 0$ can be naturally achieved if 
\begin{equation}
v_E = -\frac{3 M_1}{2\alpha_1}, \;\;v^2_{\nu}=-\frac{\alpha_5 u^2}{3\alpha_4}
\end{equation}

After achieving the desired vacuum alignment of the flavon fields responsible for generating charged lepton mass matrix and $3\times 3$ block of the light neutrino mass matrix, we focus on the $Z^{\prime}_3$ sector fields which generate the non-trivial active-sterile and sterile-sterile sectors. Since the fields in the two $Z_3$ sectors remain decoupled at renormalisable level, one can perform the analysis for respective vacuum alignments independently. Clearly from the field content shown in table \ref{table1}, the flavon and driving fields from the two $Z_3$ sectors can couple through superpotential terms suppressed by at least the third power of cut-off scale $\Lambda$ and hence can be safely neglected for our discussions. For the vacuum alignment purpose, here also we introduce a mirror copy each for the singlet fields $\chi, \psi, \zeta, \xi$ which help in their vacuum alignments as well as that of the triplet $\phi_S$. Incorporating the required driving fields $\phi^s_0, \chi_0, \psi_0, \zeta_0$, we can write down the superpotential involving the fields having non-trivial transformations under the additional $Z^{\prime}_3$ as 
\begin{align}
W^{\prime}_d & \supset \lambda_1 \phi^s_0 \phi_S \phi_S + \phi^s_0 \phi_S (\lambda_2 \bar{\chi} +\lambda^{\prime}_2 \bar{\psi} + \lambda^{\prime \prime}_2 \bar{\zeta}) +\phi_S \phi_S (\lambda_3 \chi_0 +\lambda^{\prime}_3 \psi_0 + \lambda^{\prime \prime}_3 \zeta_0) \nonumber \\
& + \lambda_4 \chi_0 \chi^2  +\lambda^{\prime}_4 \chi_0 \bar{\chi} \chi + \lambda^{\prime \prime }_4 \chi_0 \bar{\chi}^2 + \lambda_5 \psi_0 \psi^2+ \lambda^{\prime}_5 \psi_0 \psi \bar{\psi} + \lambda^{\prime \prime }_5 \psi_0 \bar{\psi}^2 +\lambda_6 \zeta_0 \zeta^2 \nonumber \\
& + \lambda^{\prime}_6 \zeta_0 \zeta \bar{\zeta} + \lambda^{\prime \prime }_6 \zeta_0 \bar{\zeta}^2 
\end{align}
For the above driving sector, we have the minimisation conditions as
\begin{equation}
\frac{\partial W^{\prime}_d}{\partial \phi^s_{01}} = \frac{2}{3} \lambda_1 (\phi^2_{S1}-\phi_{S2} \phi_{S3}) + \lambda_2 \phi_{S1} \bar{\chi} + \lambda^{\prime}_2 \phi_{S3} \bar{\psi} + \lambda^{\prime \prime}_2 \phi_{S2} \bar{\zeta} = 0
\end{equation}
\begin{equation}
\frac{\partial W^{\prime}_d}{\partial \phi^s_{02}} = \frac{2}{3} \lambda_1 (\phi^2_{S2}-\phi_{S1} \phi_{S3}) + \lambda_2 \phi_{S3} \bar{\chi} + \lambda^{\prime}_2 \phi_{S2} \bar{\psi} + \lambda^{\prime \prime}_2 \phi_{S1} \bar{\zeta} = 0
\end{equation}
\begin{equation}
\frac{\partial W^{\prime}_d}{\partial \phi^s_{03}} = \frac{2}{3} \lambda_1 (\phi^2_{S3}-\phi_{S1} \phi_{S2}) + \lambda_2 \phi_{S2} \bar{\chi} + \lambda^{\prime}_2 \phi_{S1} \bar{\psi} + \lambda^{\prime \prime}_2 \phi_{S3} \bar{\zeta} = 0
\end{equation}
\begin{equation}
\frac{\partial W^{\prime}_d}{\partial \chi_0} = \lambda_3 (\phi^2_{S1}+ 2  \phi_{S2} \phi_{S3}) + \lambda_4 \chi^2 + \lambda^{\prime}_4 \chi \bar{\chi} + \lambda^{\prime \prime}_4 \bar{\chi}^2= 0
\end{equation}
\begin{equation}
\frac{\partial W^{\prime}_d}{\partial \psi_0} = \lambda^{\prime}_3 (\phi^2_{S2}+ 2  \phi_{S1} \phi_{S3}) + \lambda_5 \psi^2 + \lambda^{\prime}_5 \psi \bar{\psi} + \lambda^{\prime \prime}_5 \bar{\psi}^2 = 0
\end{equation}
\begin{equation}
\frac{\partial W^{\prime}_d}{\partial \zeta_0} = \lambda^{\prime \prime}_3 (\phi^2_{S3}+ 2  \phi_{S1} \phi_{S2}) + \lambda_6 \zeta^2 + \lambda^{\prime}_6 \zeta \bar{\zeta} + \lambda^{\prime \prime}_6 \bar{\zeta}^2= 0
\end{equation}
The simplest possible non-trivial solution of the above minimisation equations is
$$\phi_{S1} = \phi_{S2} = \phi_{S3} = v_S, \; \bar{\chi}=\bar{\psi}=\bar{\zeta} = 0 $$
\begin{equation}
\chi^2 = u^2_1 = -\frac{3 \lambda_3 v^2_S}{\lambda_4}, \; \psi^2 = u^2_2 = -\frac{3 \lambda^{\prime}_3 v^2_S}{\lambda_5}, \; \zeta^2 = u^2_3 = -\frac{3 \lambda^{\prime \prime}_3 v^2_S}{\lambda_6}
\end{equation}
This is similar to the vacuum alignment solution chosen for $\phi_{\nu}, \eta, \bar{\eta}$ fields discussed above. However, such a choice results in $M_{es}=M_{\mu s} = M_{\tau s}$ leading to a $\mu-\tau$ symmetric $4\times 4$ mass matrix. To generate a $\mu-\tau$ symmetry breaking active-sterile sector, we need to find other possible solutions to the above minimisation conditions. If we choose $\phi_{S1}=v_{S1}, \phi_{S2}=v_{S2}, \phi_{S3}=0$, then the above minimisation conditions can be satisfied for
$$ \bar{\chi} = \frac{2\lambda_1(2 \lambda^{\prime}_2 v_{S1} v^3_{S2}- \lambda^{\prime \prime}_2 v^4_{S1})}{\lambda_2 (\lambda^{\prime}_2 v^3_{S2} + \lambda^{\prime \prime} v^3_{S1})}, \; \bar{\psi} =  \frac{2\lambda_1(2 \lambda^{\prime \prime}_2 v^3_{S1} v_{S2}- \lambda^{\prime}_2 v^4_{S2})}{\lambda^{\prime}_2 (\lambda^{\prime \prime}_2 v^3_{S1} + \lambda^{\prime} v^3_{S2})}, \; \bar{\zeta}=\frac{-6 \lambda_1 v^2_{S1} v^2_{S2}}{\lambda^{\prime \prime}_2 v^3_{S1} + \lambda^{\prime} v^3_{S2}} $$
\begin{equation}
\chi^2 = -\frac{\lambda_3 v^2_{S1}}{\lambda_4}, \; \psi^2 = -\frac{\lambda^{\prime}_3 v^2_{S2}}{\lambda_5}, \; \zeta^2 = -\frac{2 \lambda^{\prime \prime}_3 v_{S1} v_{S2}}{\lambda_6}
\end{equation}
This will give rise to $M_{es} \neq M_{\mu s} \neq M_{\tau s}$ which is required in order to produce the correct neutrino phenomenology as discussed above. Although one can find out other possible vacuum alignments, here we have shown one possible alignment which does not give the desired neutrino phenomenology and another which can give rise to the correct neutrino parameters including non-zero $\theta_{13}$.

\subsection{Scale of Flavour Symmetry Breaking}
Although the above discussion shows how the desired vacuum structure can be realised in order to give rise to the specific lepton mass matrices mentioned earlier, it does not specify the scale at which the flavons acquire vev's. Also, the cut-off scale of the theory $\Lambda$ remains unspecified. As discussed in the context of the Altarelli-Feruglio model \cite{A4TBM1,altarelli1}, one can make some simple estimates of the scale of flavon vev's and the cut-off scale $\Lambda$ as follows. Let us consider one of the elements of the active neutrino block to be of the order of $0.05$ eV, the typical scale of atmospheric mass splitting. Considering all the flavon vev's to be equal to $u$ we get
\begin{equation}
 a+2b/3 = \frac{u}{\Lambda} \frac{10 v^2_u}{3\Lambda} \approx 5 \times 10^{-11} \; \text{GeV}
 \label{scale1}
 \end{equation}
To have a meaningful expansion of different mass terms in the powers of $u/\Lambda$, one expects the expansion parameter to be less than unity $u/\Lambda < 1$. Using this in the above expression gives 
$$ \Lambda < \frac{10 v^2_u}{15 \times 10^{-11}} $$
The scale of $v_u$ can be determined from the relation $\sqrt{v^2_u+v^2_d} = 174 \; \text{GeV}$ where $v_d$ is the vev of the neutral component of the second Higgs doublet $H_d$. Assuming $\tan{\beta} = \frac{v_u}{v_d} \approx 1$, the above relation gives an upper bound on the cut-off scale as 
$$ \Lambda < 1 \times 10^{15} \; \text{GeV}$$
Since the scale of the flavon vev's $u$ is less than the cut-off scale $\Lambda$, the above upper bound on $\Lambda$ also acts like an upper bound $u$. One can find a lower bound on $u$ from the requirement of the perturbativity of the Yukawa couplings. The strongest constraint comes from the tau lepton Yukawa $Y_{\tau} < 4\pi$ which is related to the mass of tau lepton as $m_{\tau} = Y_{\tau} v_d \frac{u}{\Lambda}$. Assuming $\tan{\beta} \approx 1$ as before, this gives rise to 
$$ \frac{u}{\Lambda} > 0.001 $$
Combining both upper and lower bounds, we have 
\begin{equation}
0.001 < \frac{u}{\Lambda} < 1
\label{scale2}
\end{equation}
Using equations \eqref{scale1} and \eqref{scale2}, one can find the range of the cut-off scale as 
\begin{equation}
1 \times 10^{12} \; \text{GeV} < \Lambda < 1 \times 10^{15} \; \text{GeV}
\label{scale3}
\end{equation}
The scale of the flavon vev's will then be determined from the equation \eqref{scale2}, once the cut-off scale is specified in the range specified by the above equation \eqref{scale3}.
\subsection{Phenomenology of Flavon Fields}
We have included several flavon and driving fields in order to achieve the desired structure of lepton mass matrices along with a light sterile neutrino at eV scale. Apart from generating the correct lepton mass matrices, these fields can have several other interesting phenomenology that can offer a complementary probe of the model. Although a detailed investigation of such additional phenomenology is beyond the scope of this present work, here we note down some interesting possibilities that can be studied in this context. One interesting phenomenology of such a model is the enhancement of lepton flavour violating processes like $\mu \rightarrow e \gamma$ that can be probed at ongoing experiments like MEG \cite{MEG16}. In the context of generic supersymmetric $A_4$ models, such discussions on enhancement of $\mu \rightarrow e \gamma$ can be found in \cite{altarelliLFV}. As noted by the authors of \cite{altarelliLFV}, the large charged lepton correction introduced to generate non-zero $\theta_{13}$ from a TBM type light neutrino mass matrix usually appears in the non-diagonal terms of the charged lepton (slepton) mass matrix that could induce a too large branching ratio of $\mu \rightarrow e \gamma$. In our model, such a problem does not arise as we are generating non-zero $\theta_{13}$ from the sterile sector instead of the charged lepton sector. The enhancement of lepton flavour violation in such models could also show up in flavour violating decay of the standard model Higgs boson $(h)$ due to the mixing between the Higgs and the flavon fields. The CMS and ATLAS experiments of the LHC had provided some hints towards such a decay $(h \rightarrow \mu \tau)$ from their 8 TeV centre of mass energy data \cite{hmutauLHC}. In our model, such Higgs-flavon mixing can occur through higher order superpotential terms suppressed by at least the second power of the cut-off scale $\Lambda$. This is likely to generate a very small contribution to $h \rightarrow \mu \tau$ branching ratio. A renormalisable version of our model could provide more significant contributions to such observables currently being looked for at several experiments, if the new physics sector responsible for tiny neutrino masses lies around the TeV corner.

The flavon as well as the driving fields could also have very interesting implications in cosmology. For example, the authors of \cite{A4inflation} showed how the flavons as well as the driving fields in a typical supersymmetric $A_4$ model can give rise to cosmic inflation, a period of very rapid accelerated expansion in the early Universe \cite{inflation}. The $A_4$ flavons can also play the role of dark matter in the Universe, if the $A_4$ symmetry is broken in such a way that it leaves a remnant $Z_2$ symmetry unbroken, along with generating a $3\times3$ light neutrino mass matrix with $\theta_{13}=0$ \cite{discreteDM}. It can also be extended to accommodate a light sterile neutrino so that the sterile sector is responsible for generating the non-zero $\theta_{13}$. We leave a detailed study of these possibilities to a future work.
\section{Numerical Analysis}
\label{sec3}
In this section, we study the impact of $\mu-\tau$ symmetry in the $3\times 3$ active neutrino block of the light neutrino mass matrix on the neutrino parameters. Unbroken $\mu-\tau$ symmetry in the $3\times 3$ block of the light neutrino mass matrix in $3+1$ framework gives rise to additional constraints relating the neutrino parameters. In the diagonal charged lepton basis, the diagonalising matrix of the $4\times4$ neutrino mass matrix can be identified as the light neutrino mixing matrix. Such a $4\times4$ mixing matrix $U$ can be parametrised by six mixing angles and three Dirac CP phases. There are three additional Majorana CP phases if the light neutrinos are assumed to be of Majorana nature. The $4\times4$ unitary mixing matrix can be parametrised as \cite{sterilemutau}
\begin{equation}
U = R_{34} \tilde{R}_{24}\tilde{R}_{14}R_{23}\tilde{R}_{13} R_{12} P
\end{equation}
where the rotation matrices $R, \tilde{R}$ can be further parametrised as (for example $R_{34}$ and $\tilde{R}_{14}$)
\begin{eqnarray}
R_{34}&=&\begin{pmatrix}
1 & 0 & 0 & 0\\
0 & 1 & 0 & 0 \\
0 & 0 & c_{34} & s_{34} \\
0 & 0 & -s_{34} & c_{34}
\end{pmatrix}, \\
\tilde{R}_{14}&=&\begin{pmatrix}
c_{14} & 0 & 0 & s_{14} e^{-i\delta_{14}}\\
0 & 1 & 0 & 0 \\
0 & 0 & 1 & 0 \\
-s_{14} e^{i\delta_{14}} & 0 & 0 & c_{14}
\end{pmatrix},
\end{eqnarray}
with $c_{ij} = \cos{\theta_{ij}}, \; s_{ij} = \sin{\theta_{ij}}$ and $\delta_{ij}$ being the Dirac CP phases whereas
$$P = \text{diag} (1, e^{-i\alpha/2}, e^{-i(\beta/2 - \delta_{13})}, e^{-i(\gamma/2 - \delta_{14})})$$
contains the three Majorana CP phases. Using this form of mixing matrix, the $4\times4$ light neutrino mass matrix can be written as
\begin{equation}
M_{\nu} = U M^{\text{diag}}_{\nu} U^T,
\label{mnu}
\end{equation}
with $M^{\text{diag}}_{\nu} = \text{diag}(m_1, m_2, m_3, m_4)$ being the diagonal mass matrix. Using two mass squared differences from three neutrino global fit data \cite{schwetz14, valle14}, and another from sterile neutrino global fits \cite{globalfit}, one can write down the light neutrino mass eigenvalues in terms of the lightest neutrino mass. For NH of active neutrinos the heavier neutrino masses can be written as
$$ m_2 = \sqrt{m^2_1+\Delta m_{21}^2}, \; m_3=\sqrt{m_1^2+\Delta m_{31}^2}, \; m_4=\sqrt{m_1^2+\Delta m_{41}^2}$$
Similarly for IH of active neutrinos, the heavier masses can be written as
$$ m_1 = \sqrt{m_3^2-\Delta m_{32}^2-\Delta m_{21}^2}, \; m_2=\sqrt{m_3^2-\Delta m_{32}^2}, \; m_4=\sqrt{m_3^2+\Delta m_{43}^2}$$
where the lightest neutrino mass is $m_{\text{lightest}} = m_1$ for NH and $m_{\text{lightest}} = m_3$ for IH. Using all these, we can write down the $4\times4$ neutrino mass matrix in terms of sixteen independent parameters: four mass eigenvalues, six mixing angles, three Dirac CP phases and three Majorana CP phases. The analytical expressions of the elements of this mass matrix are given in Appendix \ref{appen1}. By demanding the active neutrino block to preserve this discrete symmetry, we numerically evaluate the neutrino parameters for two different cases, as discussed below.

\begin{figure*}
\begin{center}
\includegraphics[width=0.45\textwidth]{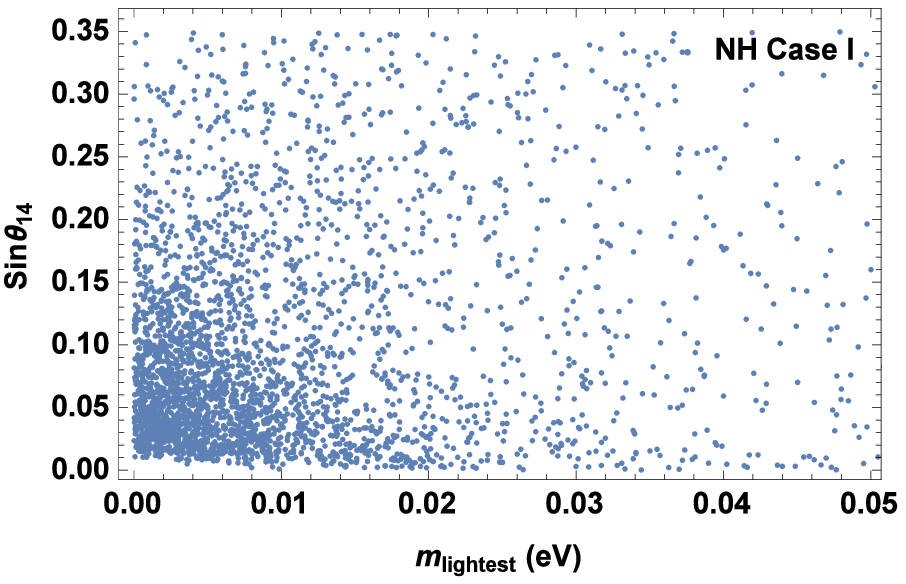}
\includegraphics[width=0.45\textwidth]{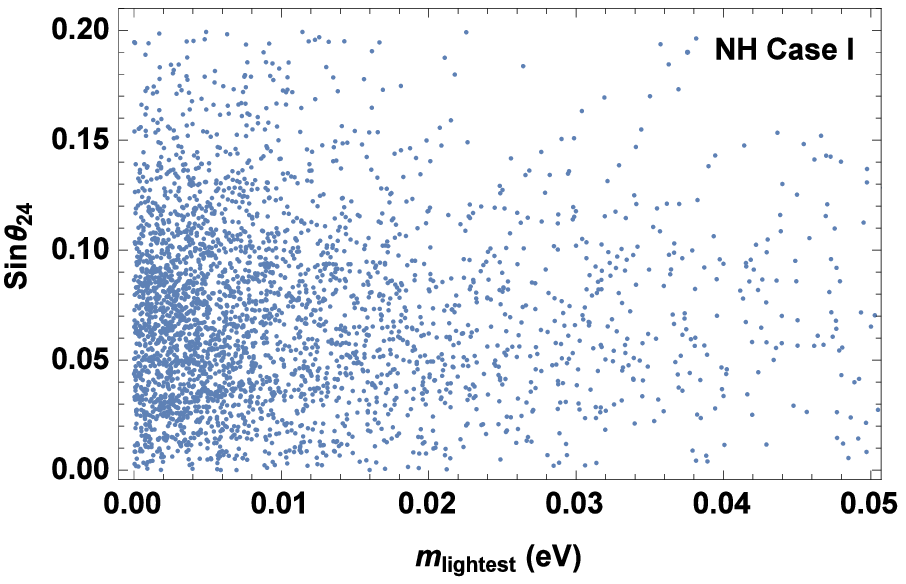} \\
\includegraphics[width=0.45\textwidth]{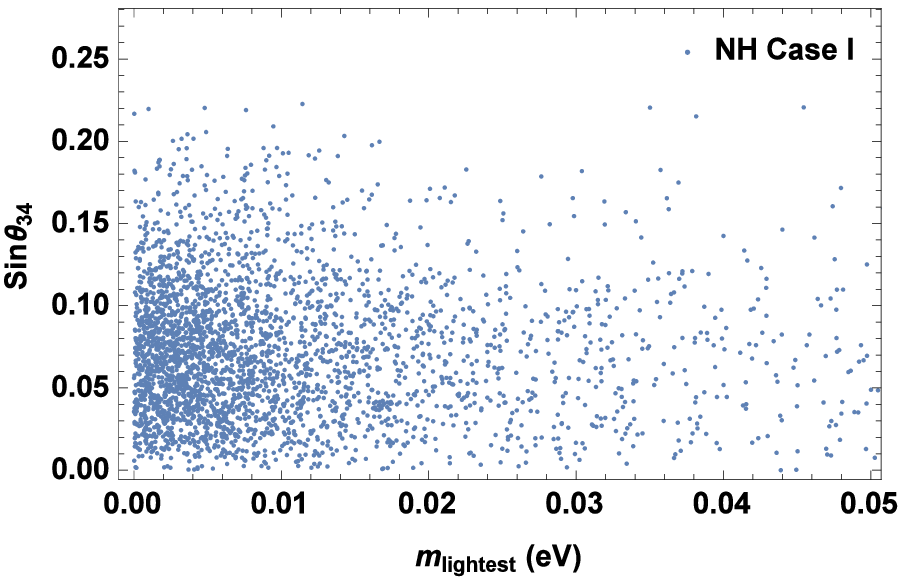}
\includegraphics[width=0.45\textwidth]{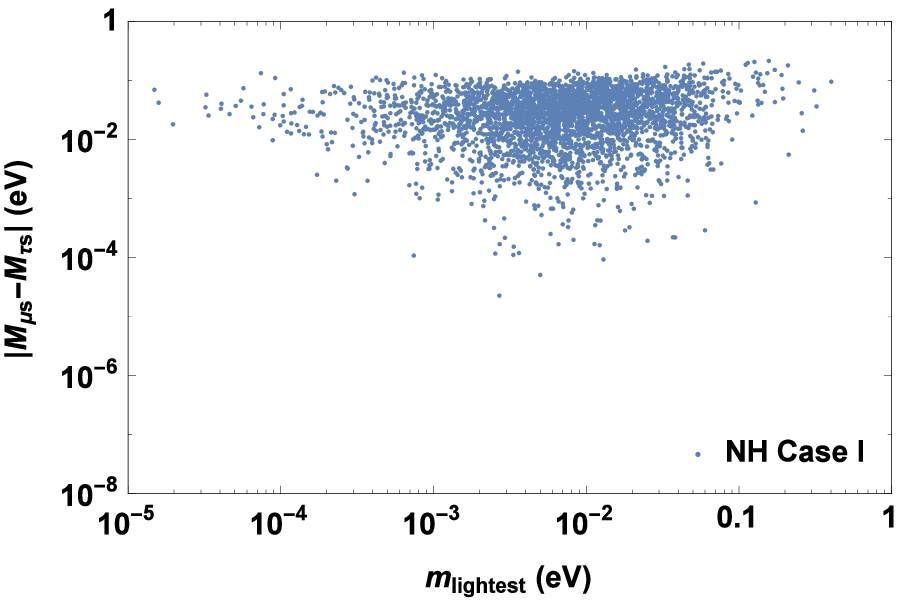}

\end{center}
\begin{center}
\caption{Active-sterile mixing angles and the amount of $\mu-\tau$ symmetry breaking in the sterile neutrino sector required to generate correct neutrino oscillation data are shown against the lightest neutrino mass for normal hierarchy and general $\mu-\tau$ symmetric $3\times3$ active neutrino block of the $4\times4$ light neutrino mass matrix.}
\label{fig1}
\end{center}
\end{figure*}

 \begin{figure*}
\begin{center}
\includegraphics[width=0.45\textwidth]{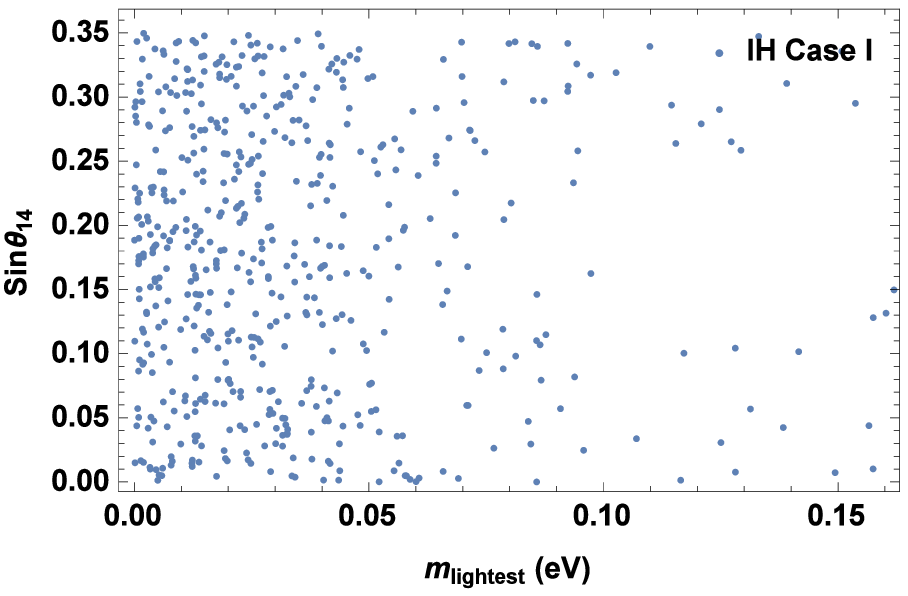}
\includegraphics[width=0.45\textwidth]{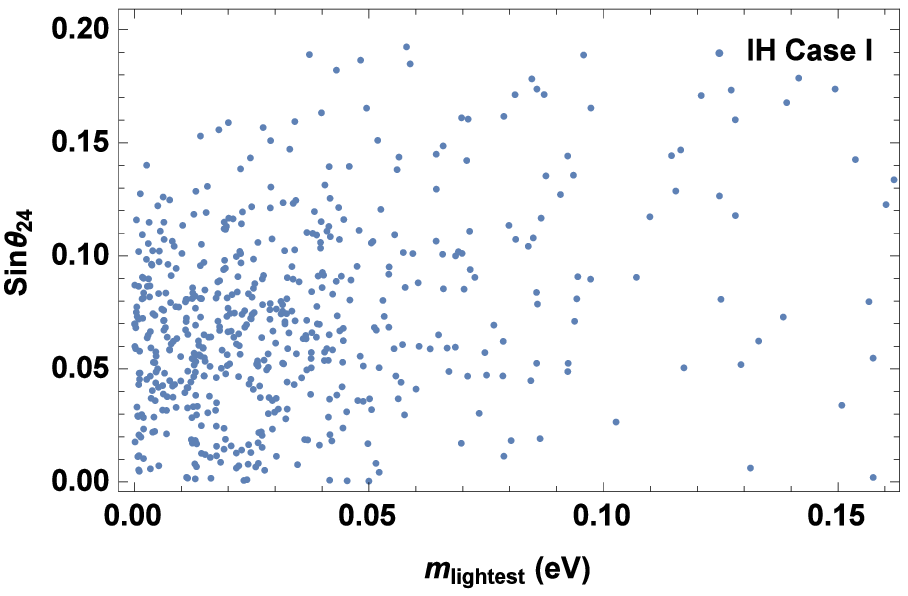} \\
\includegraphics[width=0.45\textwidth]{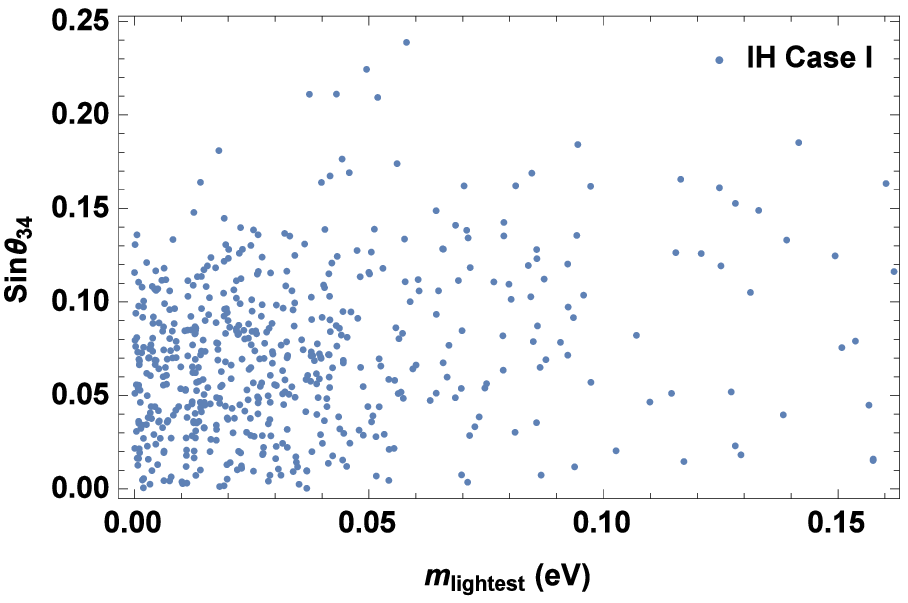}
\includegraphics[width=0.45\textwidth]{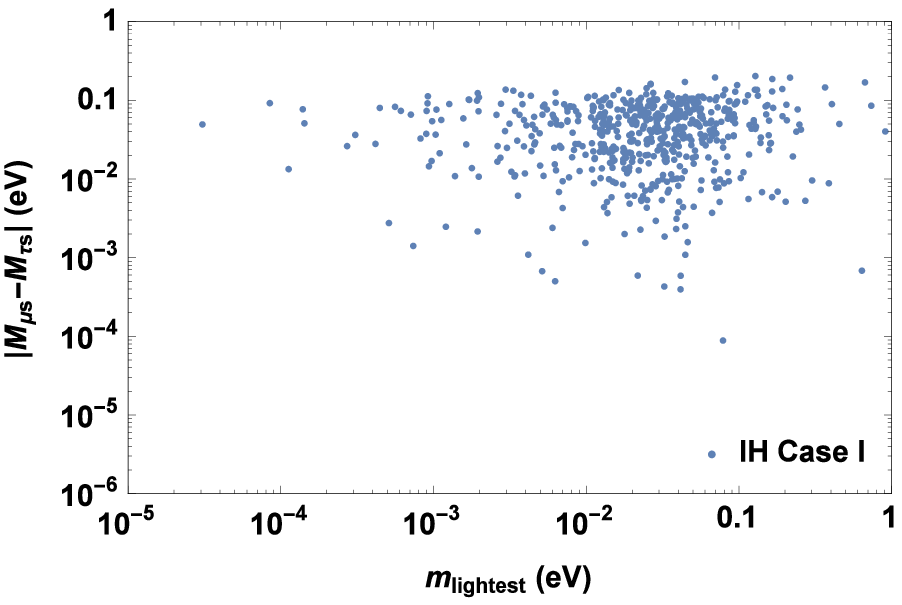}

\end{center}
\begin{center}
\caption{Active-sterile mixing angles and the amount of $\mu-\tau$ symmetry breaking in the sterile neutrino sector required to generate correct neutrino oscillation data are shown against the lightest neutrino mass for inverted hierarchy and general $\mu-\tau$ symmetric $3\times3$ active neutrino block of the $4\times4$ light neutrino mass matrix.}
\label{fig2}
\end{center}
\end{figure*}

 \begin{figure*}
\begin{center}
\includegraphics[width=0.45\textwidth]{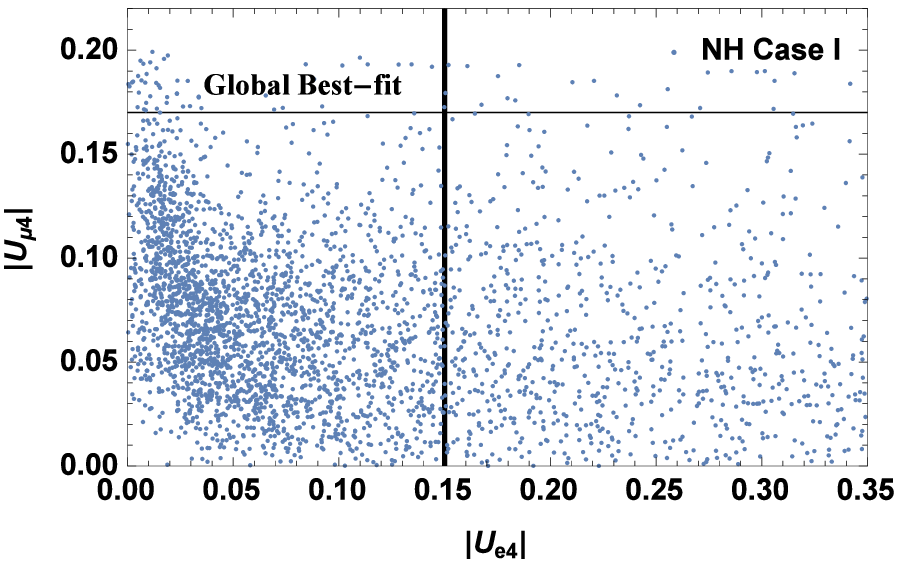}
\includegraphics[width=0.45\textwidth]{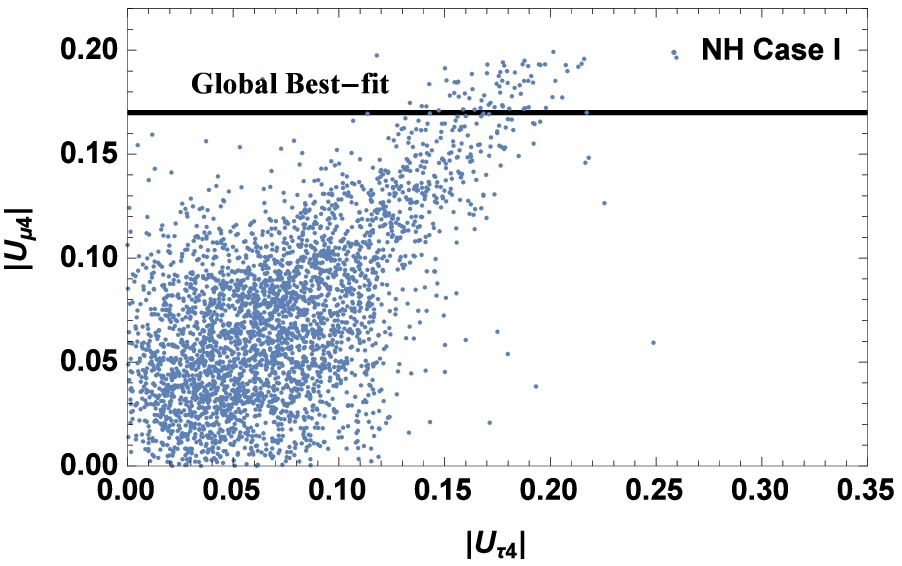} \\
\includegraphics[width=0.45\textwidth]{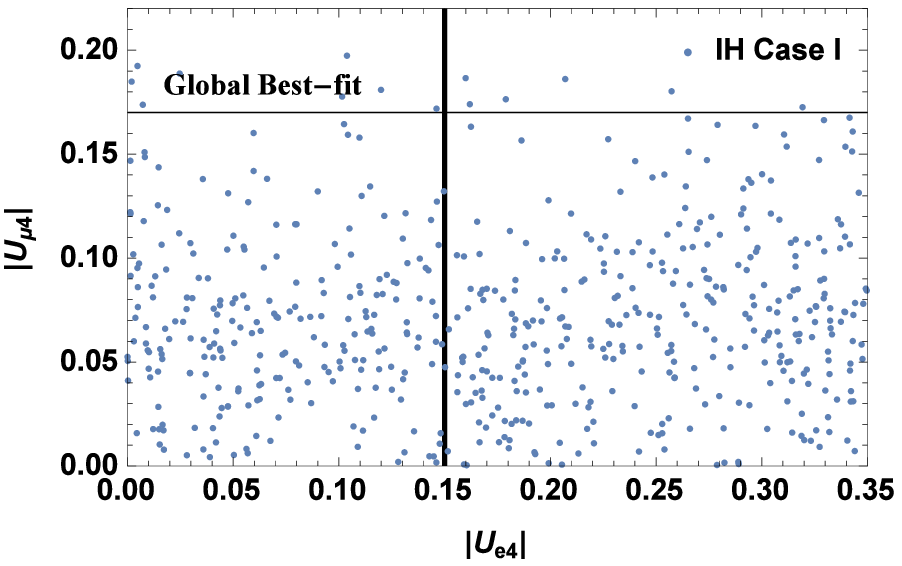}
\includegraphics[width=0.45\textwidth]{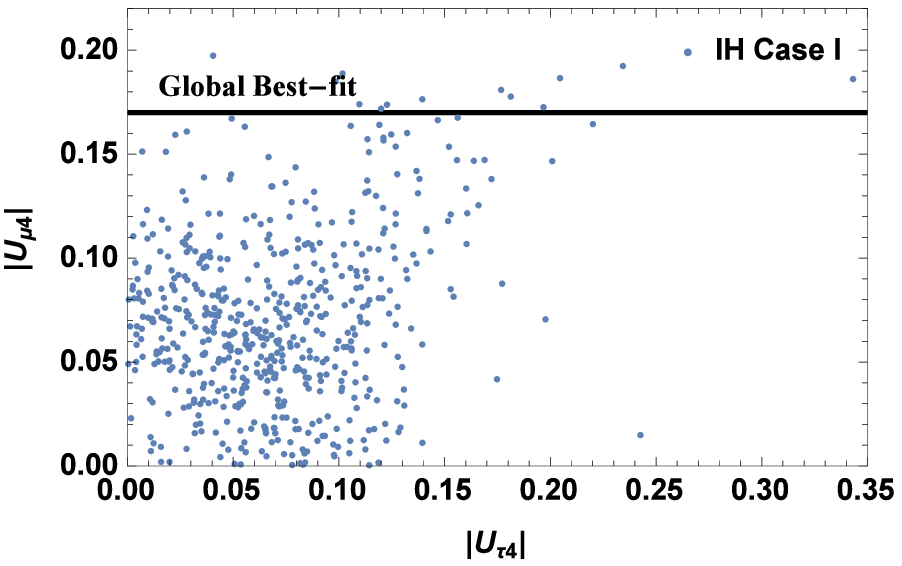}

\end{center}
\begin{center}
\caption{Active-sterile mixing allowed by general $\mu-\tau$ symmetric $3\times3$ active neutrino block of the $4\times4$ light neutrino mass matrix. The black solid lines correspond to the Global best fit values appeared in \cite{globalfit}.}
\label{fig21}
\end{center}
\end{figure*}

\begin{figure*}
\begin{center}
\includegraphics[width=0.45\textwidth]{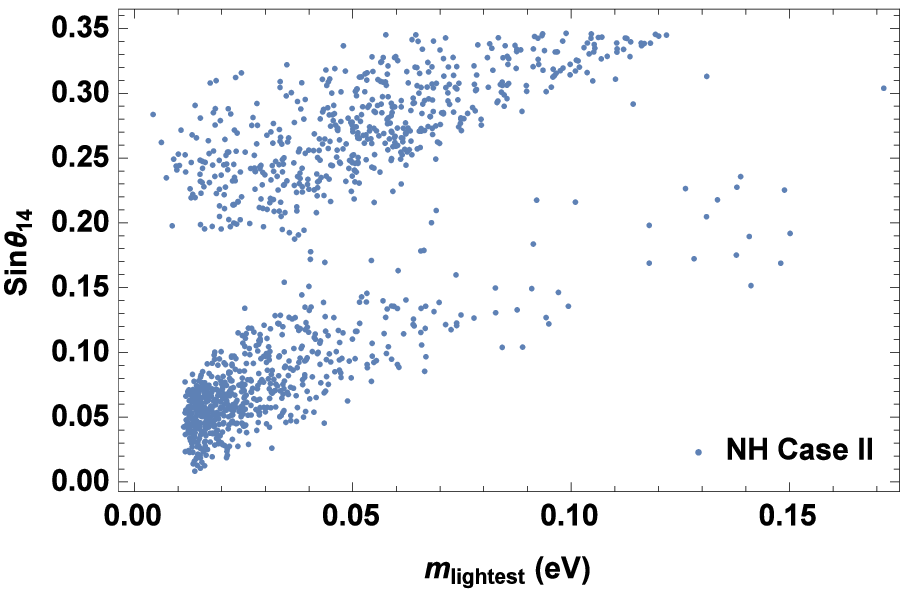}
\includegraphics[width=0.45\textwidth]{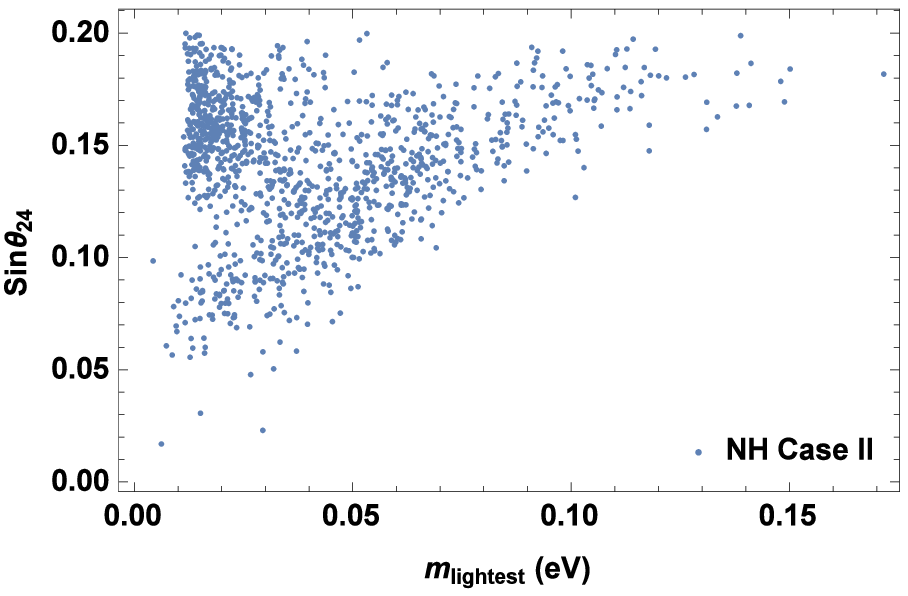} \\
\includegraphics[width=0.45\textwidth]{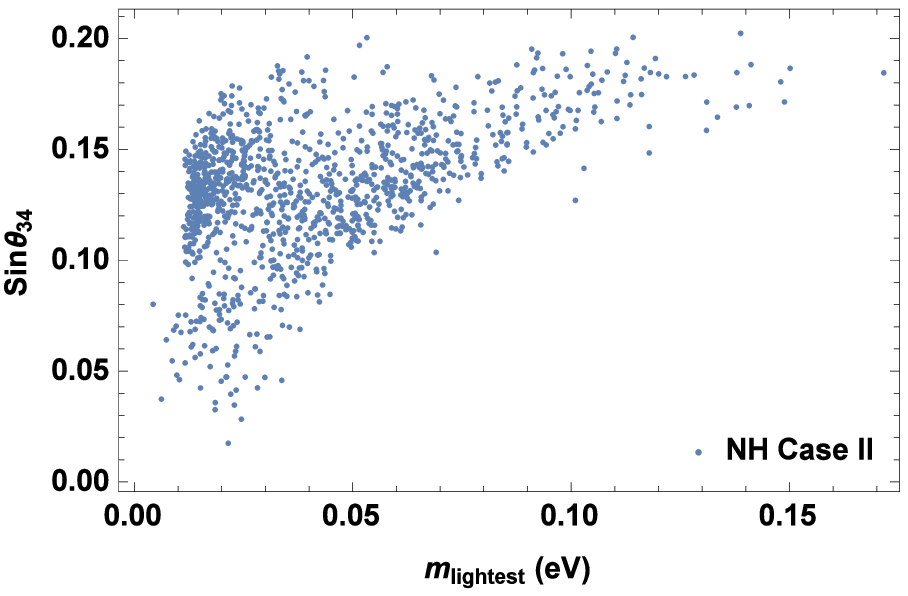}
\includegraphics[width=0.45\textwidth]{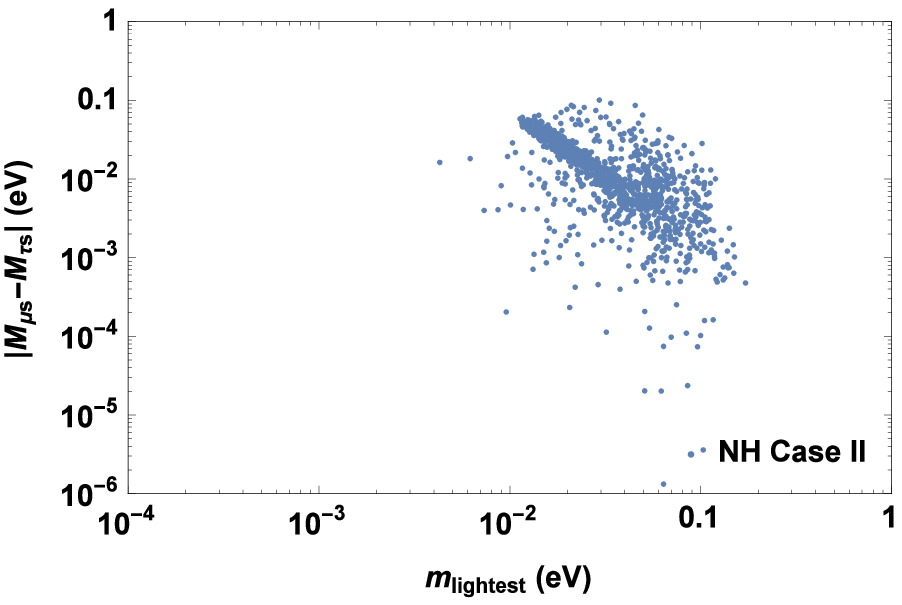}

\end{center}
\begin{center}
\caption{Active-sterile mixing angles and the amount of $\mu-\tau$ symmetry breaking in the sterile neutrino sector required to generate correct neutrino oscillation data are shown against the lightest neutrino mass for normal hierarchy and minimal $A_4$ model predicted structure of $\mu-\tau$ symmetric $3\times3$ active neutrino block of the $4\times4$ light neutrino mass matrix.}
\label{fig3}
\end{center}
\end{figure*}

 \begin{figure*}
\begin{center}
\includegraphics[width=0.45\textwidth]{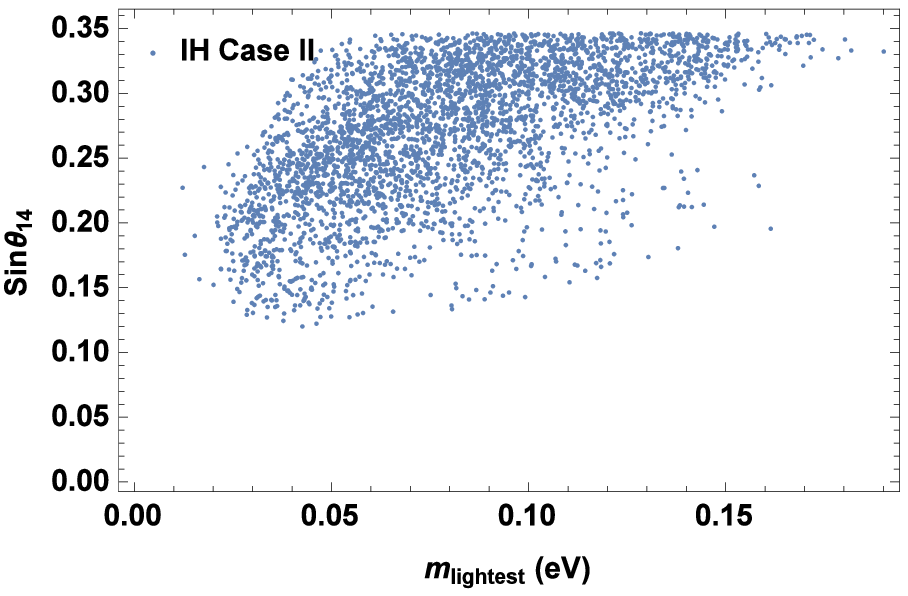}
\includegraphics[width=0.45\textwidth]{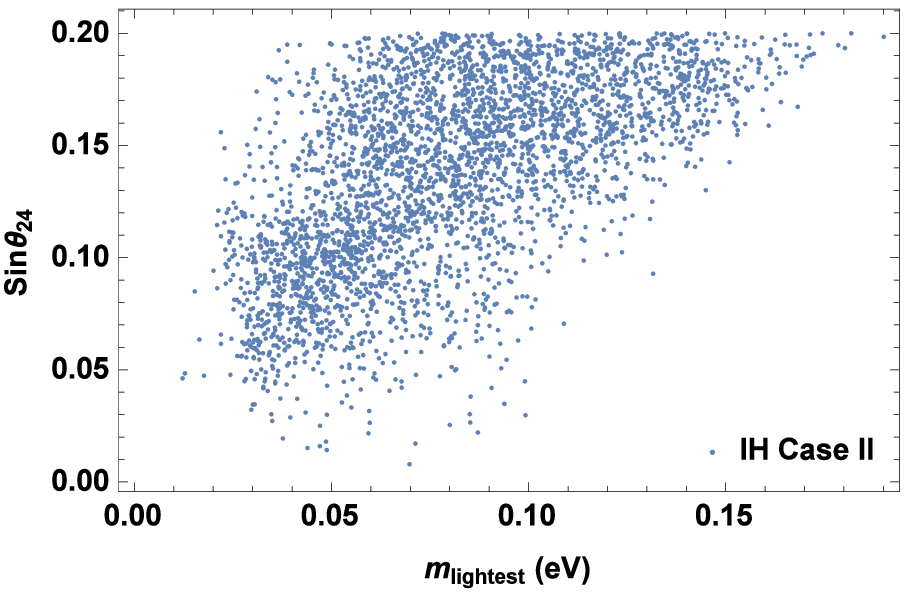} \\
\includegraphics[width=0.45\textwidth]{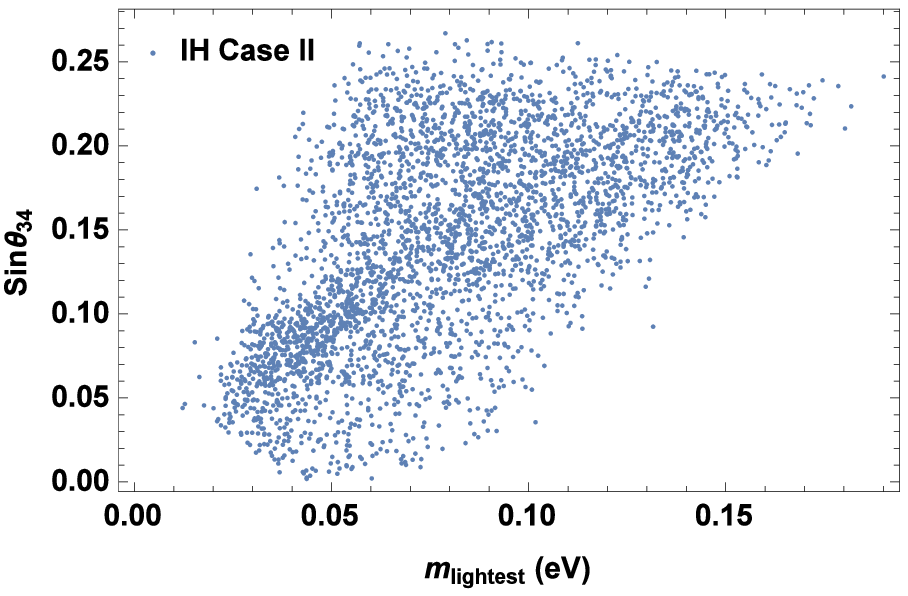}
\includegraphics[width=0.45\textwidth]{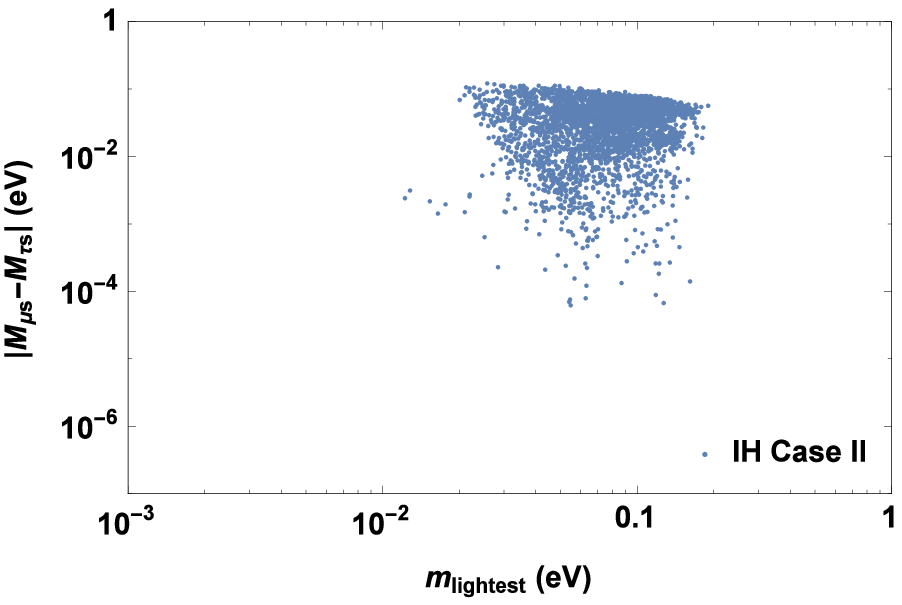}

\end{center}
\begin{center}
\caption{Active-sterile mixing angles and the amount of $\mu-\tau$ symmetry breaking in the sterile neutrino sector required to generate correct neutrino oscillation data are shown against the lightest neutrino mass for inverted hierarchy and minimal $A_4$ model predicted structure of $\mu-\tau$ symmetric $3\times3$ active neutrino block of the $4\times4$ light neutrino mass matrix.}
\label{fig4}
\end{center}
\end{figure*}

 \begin{figure*}
\begin{center}
\includegraphics[width=0.45\textwidth]{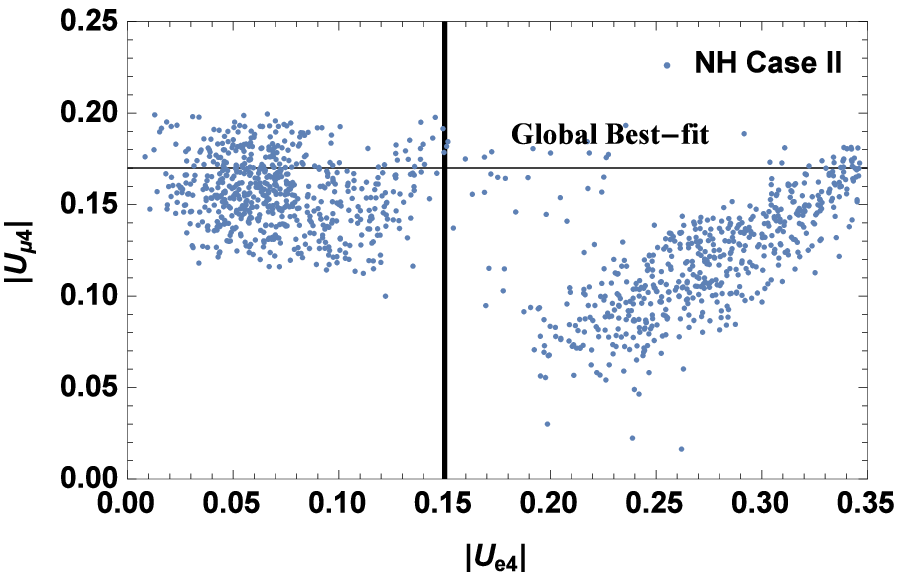}
\includegraphics[width=0.45\textwidth]{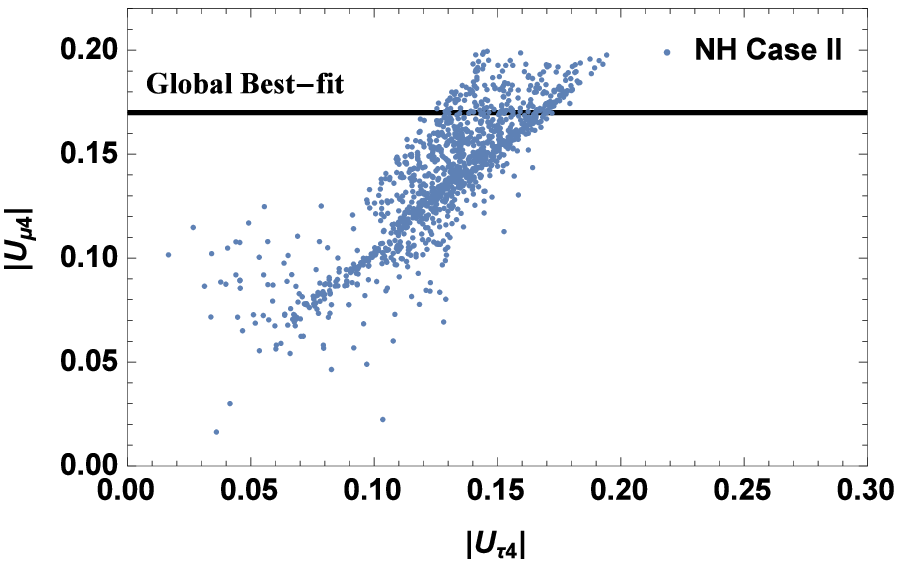} \\
\includegraphics[width=0.45\textwidth]{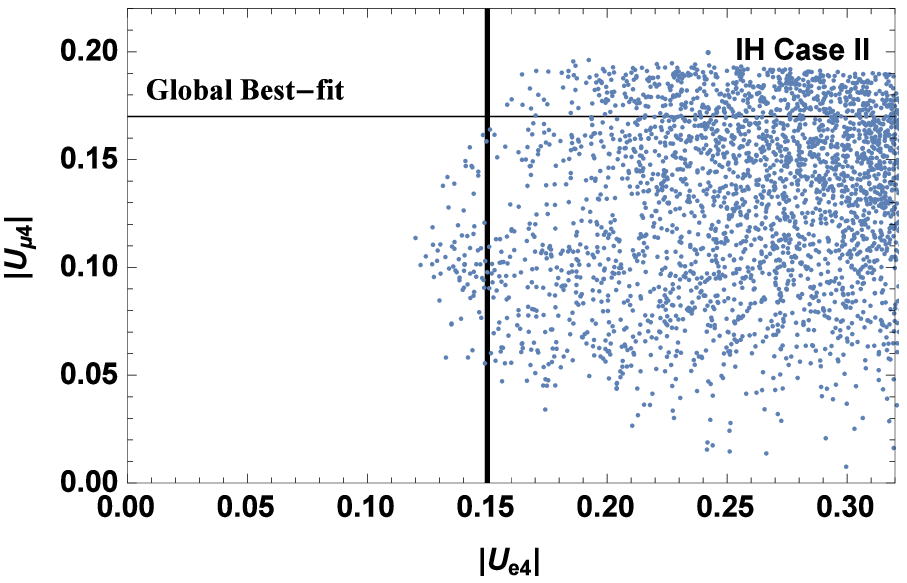}
\includegraphics[width=0.45\textwidth]{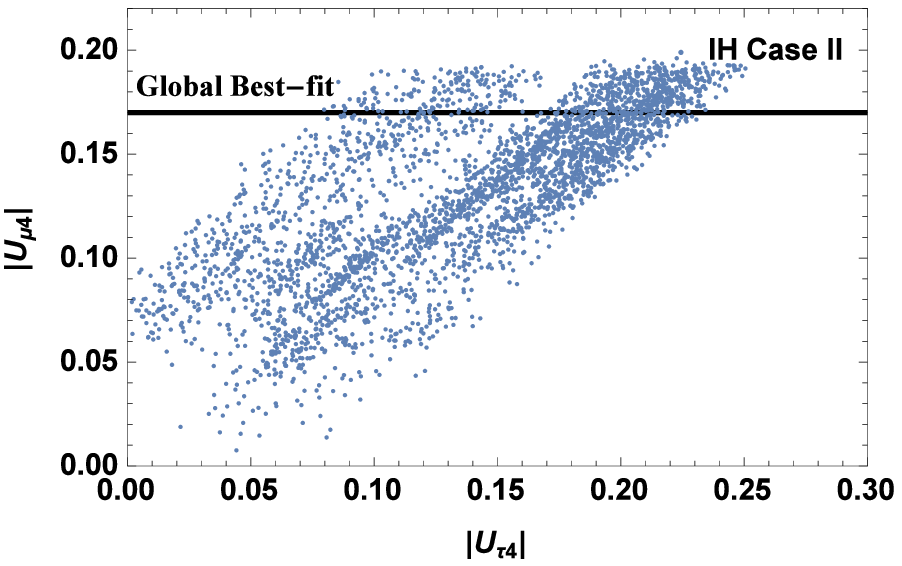}

\end{center}
\begin{center}
\caption{Active-sterile mixing allowed by the minimal $A_4$ model predicted structure of $\mu-\tau$ symmetric $3\times3$ active neutrino block of the $4\times4$ light neutrino mass matrix. The black solid lines correspond to the Global best fit values appeared in \cite{globalfit}.}
\label{fig41}
\end{center}
\end{figure*}

 \begin{figure*}
\begin{center}
\includegraphics[width=0.45\textwidth]{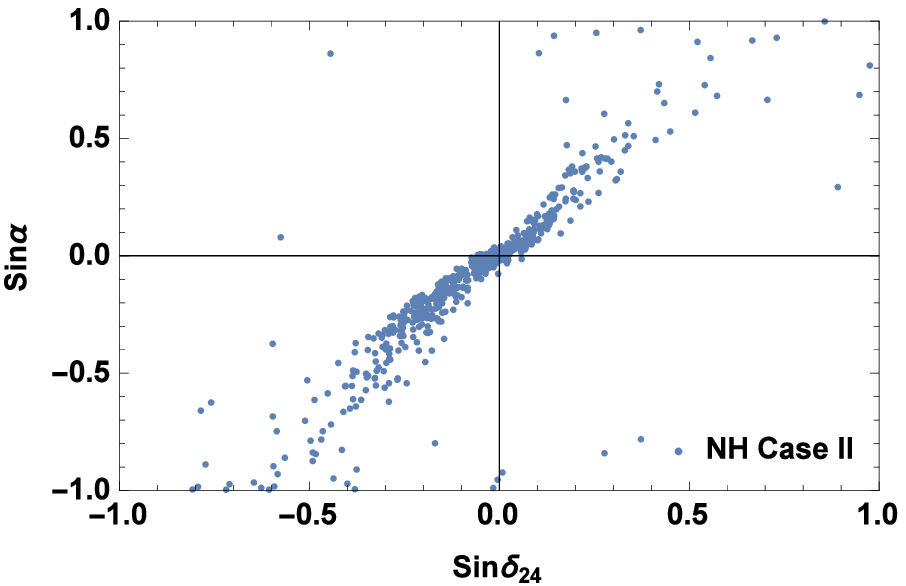}
\includegraphics[width=0.45\textwidth]{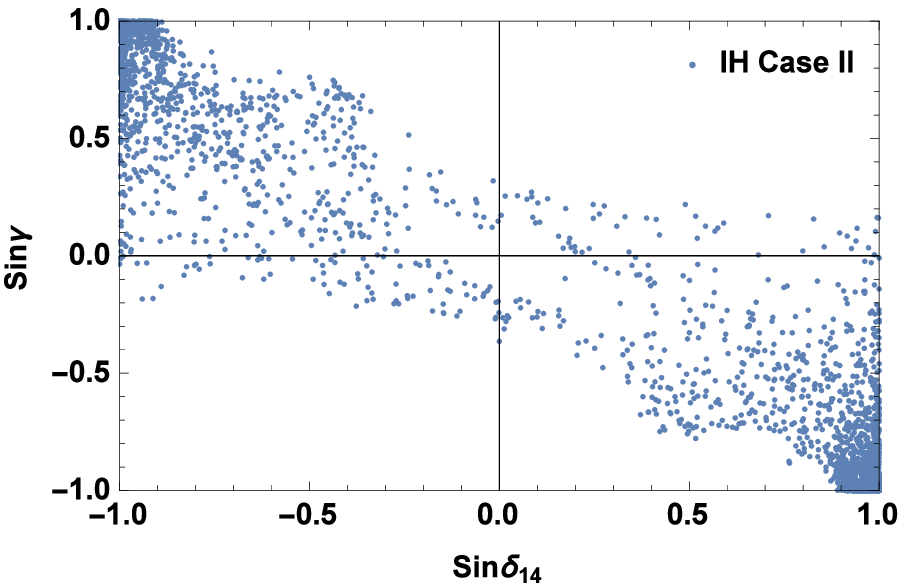}
\end{center}
\begin{center}
\caption{Correlation plots in the minimal $A_4$ model with $\mu-\tau$ symmetric $3\times3$ active neutrino block of the $4\times4$ light neutrino mass matrix.}
\label{fig5}
\end{center}
\end{figure*}

 \begin{figure*}
\begin{center}
\includegraphics[width=0.45\textwidth]{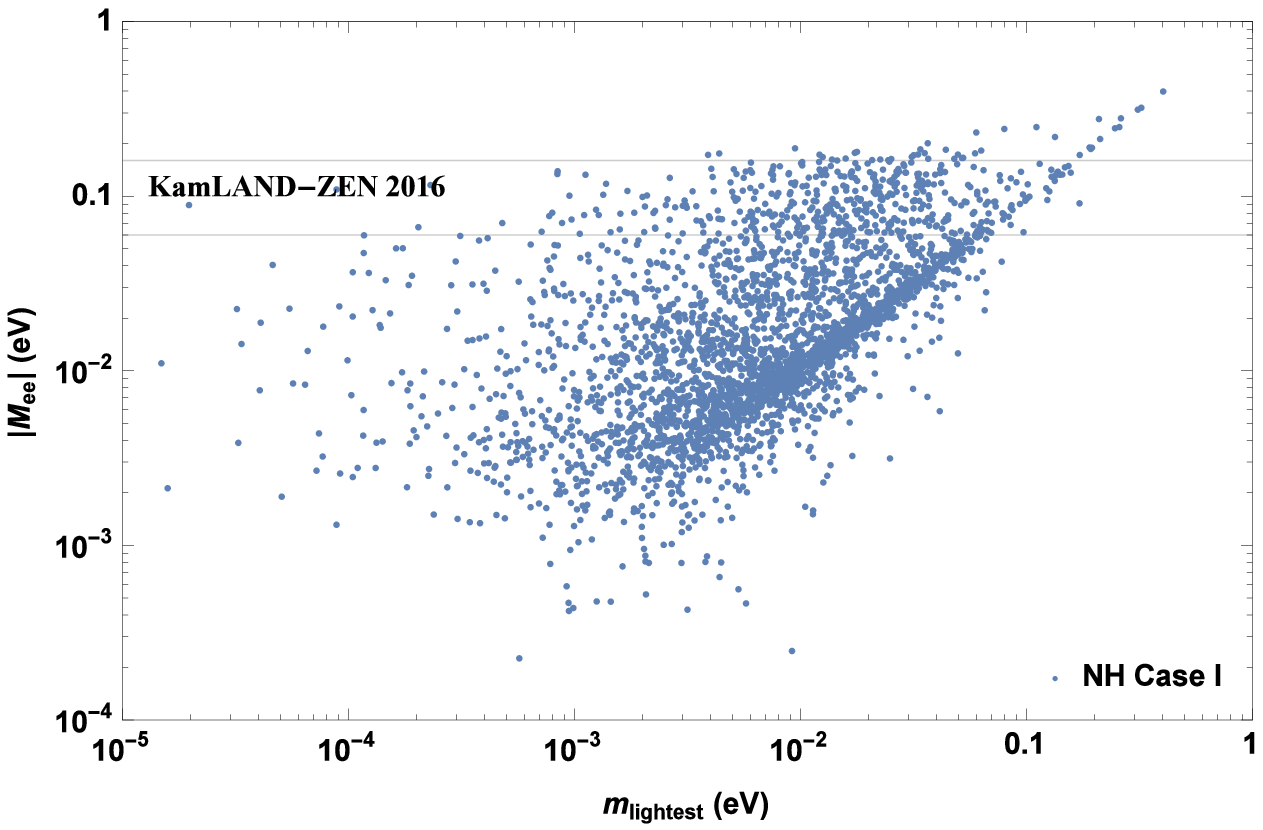}
\includegraphics[width=0.45\textwidth]{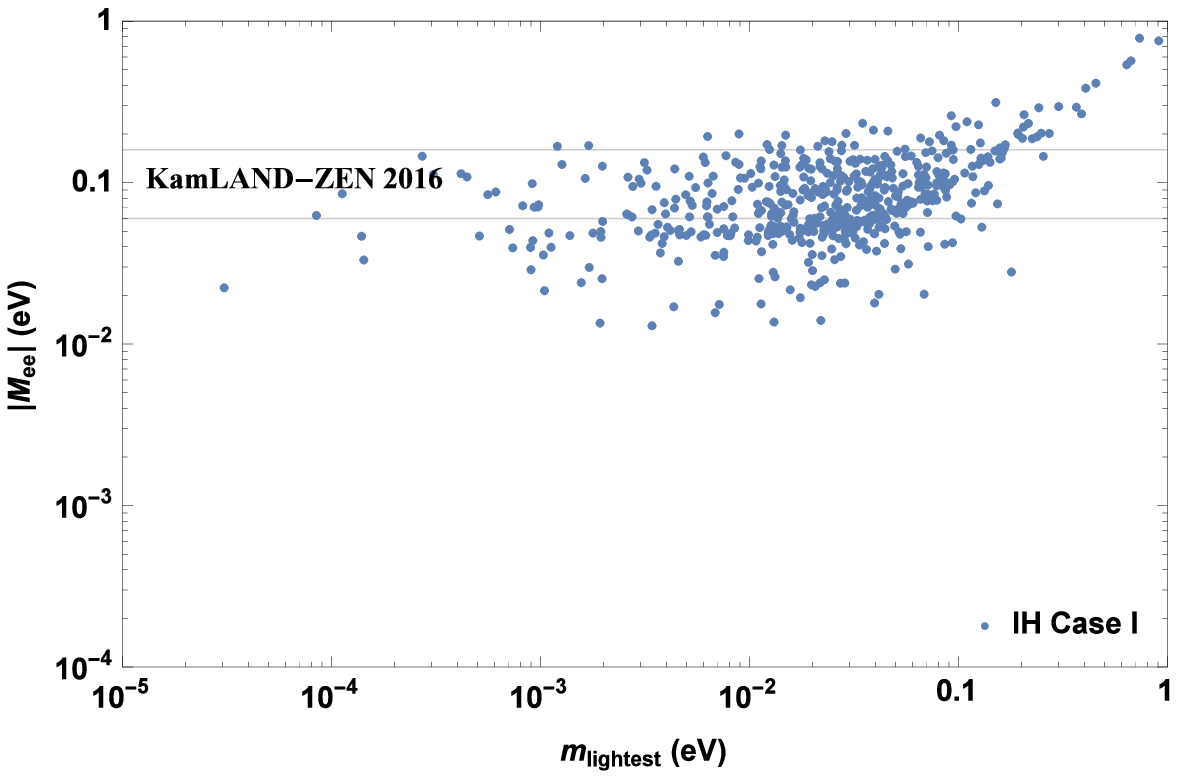} \\
\includegraphics[width=0.45\textwidth]{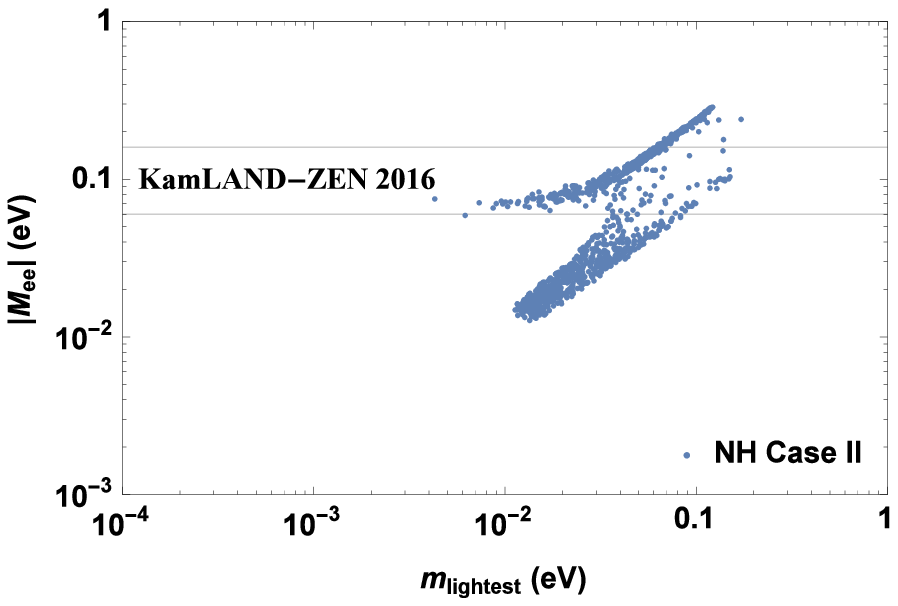}
\includegraphics[width=0.45\textwidth]{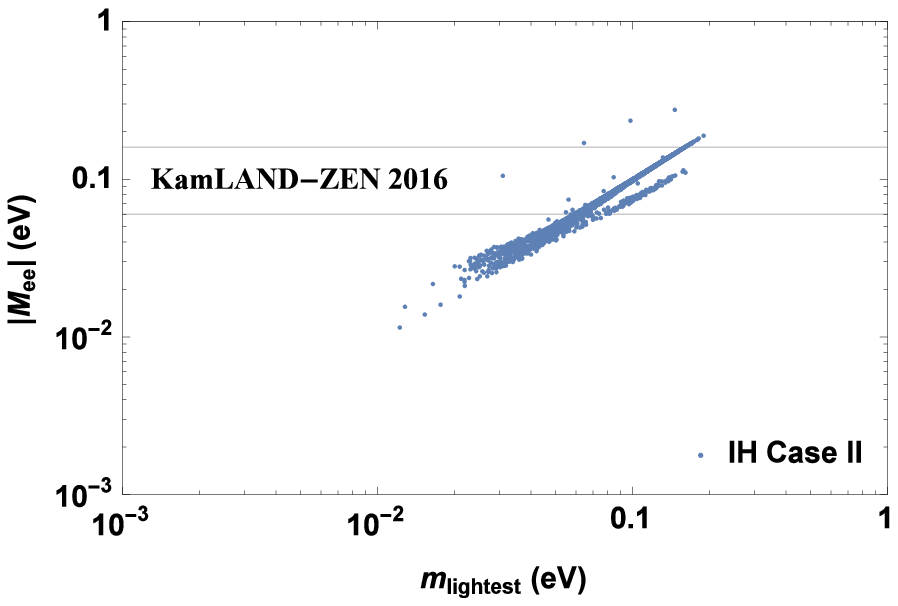}
\end{center}
\begin{center}
\caption{Contribution to effective neutrino mass $M_{ee}$ relevant for $0\nu \beta \beta$ from the $3+1$ neutrino parameter space that allows $\mu-\tau$ symmetric $3\times3$ active neutrino block in a general $4\times4$ light neutrino mass matrix.}
\label{fig6}
\end{center}
\end{figure*}
\subsection{Case I: General $\mu-\tau$ Symmetry}
For the case of general $\mu-\tau$ symmetry in the $3\times3$ block of the $4\times4$ mass matrix, one has the following constraints
$$ M_{e \mu} = M_{e \tau}, \; M_{\mu \mu} = M_{\tau \tau}$$
This is clear from the structure of the mass matrix given in equation \eqref{mutaugeneral}. Since the elements of the mass matrix are in general complex quantities, the above two constraints give rise to four real equations. These four coupled equations can be solved simultaneously to identify the $3+1$ neutrino parameter space that respects this general $\mu-\tau$ symmetric $3\times3$ block of the $4\times4$ mass matrix. We use the global fit $3\sigma$ values three active neutrino mixing angles, two mass squared differences from \cite{schwetz14, valle14}, the active-sterile mass squared difference $\Delta m_{41}^2$ (NH), $\Delta m_{43}^2$ (IH) from \cite{globalfit}. Since the leptonic CP phases are not known yet, we vary them in their $3\sigma$ allowed range $(-\pi, \pi)$. This leaves us with three active-sterile mixing angles $(\theta_{14}, \theta_{24}, \theta_{34})$ and the lightest neutrino mass $m_{\text{lightest}}=m_1 (\text{NH}), m_3 (\text{IH})$ as unknown parameters. They can be evaluated numerically by solving the four constraint equations mentioned above.
\subsection{Case II: $\mu-\tau$ Symmetry in Minimal $A_4$ Model}
For the light neutrino mass matrix with $\mu-\tau$ symmetry in the $3\times3$ block discussed within a minimal $A_4$ flavour model in section \ref{sec2}, it is straightforward to see that there are more constraints relating the mass matrix elements than in the general case discussed above. These constraints are
$$ M_{e \mu} = M_{e \tau}, \; M_{\mu \mu} = M_{\tau \tau}, \; 2 M_{e \mu}=-M_{\mu \mu}, \; M_{\mu \tau}-3 M_{e \mu} = M_{ee} $$
It is straightforward to realise these constraint equations from the structure of the mass matrix given in equation \eqref{mutauA4}. These four complex constraints give rise to eight real constraint equations relating the sixteen neutrino parameters within a $3+1$ framework discussed above. We use $3\sigma$ global fit values of five mixing angles $\theta_{12}, \theta_{23}, \theta_{13}, \theta_{14}, \theta_{24}$ and three mass squared differences leaving eight unknown parameters: six CP phases, one mixing angle $\theta_{34}$ and the lightest neutrino mass $m_{\text{lightest}}$. These eight parameters are determined by solving the eight constraints mentioned above.

\subsection{Implications for $0\nu \beta \beta$}
If neutrinos are Majorana fermions, lepton number is violated giving rise to a non-zero amplitude of neutrinoless double beta decay \footnote{It is also possible to have vanishing amplitude for $0\nu \beta \beta$ due to the interplay of different phases, even if the neutrinos are Majorana fermions}. $0\nu\beta \beta$ is a process where a heavier nucleus decays into a lighter one and two electrons $(A, Z) \rightarrow (A, Z+2) + 2e^- $ without any neutrinos in the final state. For a review on $0\nu\beta \beta$, please refer to \cite{NDBDrev}. The contribution of light neutrinos in the SM to the amplitude of $0\nu\beta \beta$ remain suppressed much below the sensitivity of ongoing experiments \cite{kamland, GERDA, kamland2} unless the lightest neutrino mass falls in the quasi-degenerate regime, which is already in tension with the Planck upper bound on the sum of absolute neutrino masses \cite{Planck15}. In the presence of a light sterile neutrino with non-trivial mixing with active neutrinos, one can have sizeable contributions to the $0\nu\beta \beta$ amplitude even if the lightest neutrino mass is much smaller than the quasi degenerate limit. Some earlier works on light sterile neutrino contributions to $0\nu\beta \beta$ can be found in \cite{ndbdsterile}.

The $0\nu\beta \beta$ amplitude corresponding to the light neutrino contribution can be written as
\begin{equation}
A_{\nu L L} \propto G^2_F \sum_i \frac{m_i U^2_{ei}}{p^2}
\end{equation}
with $p \approx 100$ MeV being the average momentum exchange for the process. In the above expression, $m_i$ are the masses of light neutrinos for $i=1,2,3,4$ whereas $G_F = 1.17 \times 10^{-5} \; \text{GeV}^{-2}$ is the Fermi coupling constant and $U$ is the light neutrino mixing matrix. Thus, the light neutrino contribution can be written in terms of $\lvert M_{ee} \rvert = \lvert U^2_{ei} m_i \rvert$ which is a function of mixing angles $\theta_{12}, \theta_{13}, \theta_{14}$, Majorana CP phases $\alpha, \beta, \gamma$ and four mass eigenvalues $m_{1,2,3,4}$ as shown in appendix \ref{appen1}. Using the numerically evaluated neutrino parameters for the two cases discussed above, one can calculate the numerical value of $\lvert M_{ee} \rvert$ and compare against the most recent KamLAND-Zen results,
according to which the upper bound on $ \lvert M_{ee} \rvert $ is (0.06 - 0.16) eV at $90\%$ C.L. \cite{kamland2}.

\section{Results and Conclusion}
\label{sec4}
We have studied the possibility of generating non-zero $\theta_{13}$ by breaking $\mu-\tau$ symmetry only in the sterile neutrino sector, while keeping it unbroken in the $3\times3$ active neutrino mass matrix. In a scenario with three active and one light sterile neutrino, the $4\times4$ mass matrix with $\mu-\tau$ symmetric $3\times3$ active neutrino block can give rise to correct neutrino oscillation data provided the $\mu-\tau$ symmetry is broken in the sterile neutrino sector due to the inequality $M_{\mu s} \neq M_{\tau s}$. We have proposed a supersymmetric model based on $A_4 \times Z_3 \times Z^{\prime}_3$ flavour symmetry that can give rise to the desired $4\times 4$ light neutrino mass matrix. We also discuss in details, the possible vacuum alignment of the flavon fields that can generate a active-sterile sector which breaks the $\mu-\tau$ symmetry and hence can give rise to non-zero $\theta_{13}$. Considering such a light neutrino mass matrix to have a general structure (with $\mu-\tau$ symmetric active neutrino block and $\mu-\tau$ breaking active-sterile sector) as well as the one within a minimal $A_4$ flavour model, we numerically evaluate the full parameter space that can give rise to such a mass matrix. In the general case (denoted as case I in the previous section), the three active-sterile mixing angles generated from the constraint equations along with the lightest neutrino mass are shown in figures \ref{fig1}, \ref{fig2} for NH and IH respectively. It is interesting to note from the first three panels of these two figures that even if the present experiments \cite{icecube1, sterileExpt16} completely rule out the light sterile neutrino parameter space suggested by \cite{globalfit} in order to explain the neutrino anomalies discussed before, the present framework of generating non-zero reactor mixing angle can still survive. This is due to the fact that very small values of active-sterile mixing angles (not excluded by present experiments) are also consistent with correct neutrino oscillation data in the active sector. The last panel of these two figures \ref{fig1}, \ref{fig2} show the extent of $\mu-\tau$ symmetry breaking $\lvert \Delta M \rvert = \lvert M_{\tau s} - M_{\mu s} \rvert$ required to generate correct neutrino oscillation data. We then show the corresponding active-sterile mixing elements of the $4\times4$ mixing matrix $U$ in figure \ref{fig21} and compare them with their best fit values $\lvert U_{e4} \rvert = 0.15, \lvert U_{\mu 4} = 0.17$ that appeared in \cite{globalfit}. It can be seen that the requirement of producing the correct active neutrino oscillation data is still consistent with active-sterile mixing elements much smaller than the present global fit values. We repeat the same calculation for case II that is, the minimal $A_4 \times Z_3 \times Z^{\prime}_3$ model discussed above. The plots in figures \ref{fig3}, \ref{fig4} show the relevant parameter space in active-sterile mixing angles, lightest neutrino mass for NH and IH respectively. Similar to the general case, the last panel of figures \ref{fig3}, \ref{fig4} show the required deviation from $\mu-\tau$ symmetry in the sterile neutrino sector. Similar to the case I, here also we show the mixing matrix elements in comparison with their best fit values in figure \ref{fig41}. Interestingly, in this case, not too many allowed active-sterile mixing elements lie in the region satisfying $\lvert U_{e4} \rvert < 0.15, \lvert U_{\mu 4} < 0.17$. Thus, if future oscillation experiments rule out $\lvert U_{e4} \rvert > 0.10, \lvert U_{\mu 4} > 0.12$, then this scenario within a minimal $A_4 \times Z_3 \times Z^{\prime}_3$ model will be ruled out completely. Figure \ref{fig5} shows some interesting correlations between the CP phases in case II.

We also calculate the effective neutrino mass $M_{ee}$ in order to check the implications of the neutrino parameters evaluated above for neutrinoless double beta decay. The corresponding values of $\lvert M_{ee} \rvert $ are shown as a function of the lightest neutrino mass $m_{\text{lightest}}$ for both case I, II as well as NH, IH in figure \ref{fig6}. It is interesting to note from these figures that the latest KamLAND-Zen bound \cite{kamland2} already rules out a small part of parameter space. An order of amplitude improvement in these experimental searches for $0\nu\beta \beta$ will in fact rule out three of the scenarios discussed in this work. Only the general $\mu-\tau$ symmetric case with NH will survive in that case, as it has some parameter space which predicts very small values of $\lvert M_{ee} \rvert $, way below the present sensitivity of KamLAND-Zen experiment.

To summarise, after proposing a flavour symmetry model for $3+1$ light neutrino scenario, we have evaluated the $3+1$ neutrino parameter space obeying a discrete $\mu-\tau$ symmetry in the $3\times3$ block of the $4\times4$ neutrino mass matrix which is in agreement with the current neutrino data. Such a symmetric active neutrino block of the $4\times4$ mass matrix restricts the neutrino parameter space to some specific values, satisfying the constraints imposed by the $\mu-\tau$ symmetry, both in general as well as in the minimal $A_4 \times Z_3 \times Z^{\prime}_3$ model. This constrained parameter space can have interesting implications at oscillation as well as neutrinoless double beta decay experiments. We found that, correct active neutrino oscillation data can be generated for a $\mu-\tau$ symmetric $3\times3$ active neutrino block within a $4\times4$ mass matrix, even if the active-sterile mixing angles are smaller than the ones required by LSND, MiniBooNE and other neutrino data showing anomalies. In the minimal $A_4$ model however, the allowed active-sterile mixing elements lie very close to the global best fit values \cite{globalfit}. This is interesting in the light of ongoing experiments \cite{icecube1, sterileExpt16} which are claiming to rule out most part of parameter space required to explain the neutrino anomalies. Irrespective of whether the neutrino oscillation experiments will be able to rule out these scenarios (where non-zero $\theta_{13}$ originates from $\mu-\tau$ symmetry breaking only in the sterile sector) in near future or not, an order of magnitude improvement in $0\nu\beta \beta$ sensitivity should be able to rule out most of these models except a general $\mu-\tau$ symmetric case with normal hierarchy as discussed above. Therefore, apart from the general $\mu-\tau$ symmetric case with normal hierarchy, all other cases discussed in this work can be ruled out or verified either in neutrino oscillation or neutrinoless double beta decay experiments or both in near future. Probing these scenarios could also shed more light into the fundamental symmetries behind the origin of leptonic mixing, similar to the particular example of $A_4 \times Z_3 \times Z^{\prime}_3$ symmetric model discussed in this work. This can also have very interesting implications for the creation of matter-antimatter asymmetry through the mechanism of leptogenesis specially due to the fact that flavour symmetric seesaw models (in 3 light neutrino picture) with exact TBM mixing implies a vanishing lepton asymmetry \cite{lepto1}. We leave a this interesting study for an upcoming work.

\appendix
\appendix
\section{$A_4$ product rules}
\label{appen2}
$A_4$, the symmetry group of a tetrahedron, is a discrete non-abelian group of even permutations of four objects. It has four irreducible representations: three one-dimensional and one three dimensional which are denoted by $\bf{1}, \bf{1'}, \bf{1''}$ and $\bf{3}$ respectively, being consistent with the sum of square of the dimensions $\sum_i n_i^2=12$. Their product rules are given as
$$ \bf{1} \otimes \bf{1} = \bf{1}$$
$$ \bf{1'}\otimes \bf{1'} = \bf{1''}$$
$$ \bf{1'} \otimes \bf{1''} = \bf{1} $$
$$ \bf{1''} \otimes \bf{1''} = \bf{1'}$$
$$ \bf{3} \otimes \bf{3} = \bf{1} \otimes \bf{1'} \otimes \bf{1''} \otimes \bf{3}_a \otimes \bf{3}_s $$
where $a$ and $s$ in the subscript corresponds to anti-symmetric and symmetric parts respectively. Denoting two triplets as $(a_1, b_1, c_1)$ and $(a_2, b_2, c_2)$ respectively, their direct product can be decomposed into the direct sum mentioned above as
$$ \bf{1} \backsim a_1a_2+b_1c_2+c_1b_2$$
$$ \bf{1'} \backsim c_1c_2+a_1b_2+b_1a_2$$
$$ \bf{1''} \backsim b_1b_2+c_1a_2+a_1c_2$$
$$\bf{3}_s \backsim (2a_1a_2-b_1c_2-c_1b_2, 2c_1c_2-a_1b_2-b_1a_2, 2b_1b_2-a_1c_2-c_1a_2)$$
$$ \bf{3}_a \backsim (b_1c_2-c_1b_2, a_1b_2-b_1a_2, c_1a_2-a_1c_2)$$

\section{Light neutrino mass matrix elements}
\label{appen1}

{\small \begin{widetext}
\begin{equation}
M_{ee} = c_{12}^2 c_{13}^2 c_{14}^2 m_1+e^{- i \alpha } c_{13}^2 c_{14}^2 m_2 s_{12}^2+e^{- i \beta } c_{14}^2 m_3 s_{13}^2+e^{-i \gamma } m_4 s_{14}^2 \nonumber
\end{equation}
\begin{eqnarray}
M_{e\mu} &=&-e^{-i \delta _{24}} c_{14} \big(e^{i \delta _{24}} c_{12} c_{13} c_{23} c_{24} \big(m_1-e^{- i \alpha } m_2\big) s_{12}-e^{i \big(\delta
_{13}+\delta _{24}\big)} c_{13} c_{24} \big(e^{- i \beta } m_3-e^{- i \alpha } m_2 s_{12}^2\big) s_{13} s_{23} \nonumber \\
 && +e^{i \big(2 \alpha +\delta _{14}\big)}M
c_{13}^2 m_2 s_{12}^2 s_{14} s_{24}-e^{i \delta _{14}} \big(e^{- i \gamma } m_4-e^{- i \beta } m_3 s_{13}^2\big) s_{14} s_{24}+c_{12}^2 c_{13}
m_1 \big(e^{i \big(\delta _{13}+\delta _{24}\big)} c_{24} s_{13} s_{23} \nonumber  \\
&& +e^{i \delta _{14}} c_{13} s_{14} s_{24}\big)\big) \nonumber
\end{eqnarray}
\begin{eqnarray}
M_{e\tau}&=&c_{14} \big(-e^{i \big(- \alpha +\delta _{14}\big)} c_{13}^2 c_{24} m_2 s_{12}^2 s_{14} s_{34}+e^{i \delta _{14}} c_{24} \big(e^{- i \gamma
} m_4-e^{- i \beta } m_3 s_{13}^2\big) s_{14} s_{34} \nonumber \\
&& +c_{12} c_{13} \big(m_1-e^{- i \alpha } m_2\big) s_{12} \big(c_{34} s_{23}+e^{i \delta
_{24}} c_{23} s_{24} s_{34}\big)+e^{i \delta _{13}} c_{13} \big(e^{- i \beta } m_3-e^{- i \alpha } m_2 s_{12}^2\big) s_{13} \big(c_{23} c_{34} \nonumber \\
&& -e^{i
\delta _{24}} s_{23} s_{24} s_{34}\big)-c_{12}^2 c_{13} m_1 \big(e^{i \delta _{13}} c_{23} c_{34} s_{13}+\big(e^{i \delta _{14}} c_{13} c_{24}
s_{14}-e^{i \big(\delta _{13} +\delta _{24}\big)} s_{13} s_{23} s_{24}\big) s_{34}\big)\big) \nonumber
\end{eqnarray}
\begin{eqnarray}
M_{\mu\mu} &=&e^{ i \big(-\gamma + 2\delta_{14}- 2 \delta_{24}\big)} c_{14}^2 m_4 s_{24}^2+e^{- i \beta } m_3 \big(e^{i \delta _{13}} c_{13} c_{24} s_{23}-e^{i
\big(\delta _{14}-\delta _{24}\big)} s_{13} s_{14} s_{24}\big){}^2+e^{- i \alpha } m_2 \big(c_{12} c_{23} c_{24}\nonumber \\
&& +s_{12} \big(-e^{i \delta
_{13}} c_{24} s_{13} s_{23}-e^{i \big(\delta _{14}-\delta _{24}\big)} c_{13} s_{14} s_{24}\big)\big){}^2+m_1 \big(c_{23} c_{24} s_{12}+c_{12}
\big(e^{i \delta _{13}} c_{24} s_{13} s_{23} \nonumber \\
&& +e^{i \big(\delta _{14}-\delta _{24}\big)} c_{13} s_{14} s_{24}\big)\big){}^2 \nonumber
\end{eqnarray}
\begin{eqnarray}
M_{\mu\tau} &=& e^{i \big(- \gamma +2 \delta _{14}-\delta _{24}\big)} c_{14}^2 c_{24} m_4 s_{24} s_{34}+e^{i \big(2 \beta +\delta _{13}\big)} m_3 \big(e^{i
\delta _{13}} c_{13} c_{24} s_{23}-e^{i \big(\delta _{14}-\delta _{24}\big)} s_{13} s_{14} s_{24}\big)  \nonumber \\
&&\big(-e^{-i \big(\delta _{13}-\delta
_{14}\big)} c_{24} s_{13} s_{14} s_{34}+c_{13} \big(c_{23} c_{34}-e^{i \delta _{24}} s_{23} s_{24} s_{34}\big)\big)+m_1 \big(-c_{23} c_{24}
s_{12}+c_{12} \big(-e^{i \delta _{13}} c_{24} s_{13} s_{23} \nonumber \\
&& -e^{i \big(\delta _{14}-\delta _{24}\big)} c_{13} s_{14} s_{24}\big)\big) \big(s_{12}
\big(c_{34} s_{23}+e^{i \delta _{24}} c_{23} s_{24} s_{34}\big)+c_{12} \big(-e^{i \delta _{14}} c_{13} c_{24} s_{14} s_{34}-e^{i \delta _{13}}
s_{13} \big(c_{23} c_{34} \nonumber \\
&&-e^{i \delta _{24}} s_{23} s_{24} s_{34}\big)\big)\big)+e^{- i \alpha } m_2 \big(c_{12} c_{23} c_{24}+s_{12} \big(-e^{i
\delta _{13}} c_{24} s_{13} s_{23}-e^{i \big(\delta _{14}-\delta _{24}\big)} c_{13} s_{14} s_{24}\big)\big) \big(-c_{12} \big(c_{34} s_{23} \nonumber \\
&&+e^{i
\delta _{24}} c_{23} s_{24} s_{34}\big)+s_{12} \big(-e^{i \delta _{14}} c_{13} c_{24} s_{14} s_{34}-e^{i \delta _{13}} s_{13} \big(c_{23} c_{34}-e^{i
\delta _{24}} s_{23} s_{24} s_{34}\big)\big)\big) \nonumber
\end{eqnarray}
\begin{eqnarray}
M_{\tau \tau} &=& e^{ i \big(-\gamma + 2 \delta_{14}\big)} c_{14}^2 c_{24}^2 m_4 s_{34}^2+e^{ i \big(-\beta +2\delta_{13}\big)} m_3 \big(e^{-i \big(\delta
_{13}-\delta _{14}\big)} c_{24} s_{13} s_{14} s_{34}+c_{13} \big(-c_{23} c_{34}+e^{i \delta _{24}} s_{23} s_{24} s_{34}\big)\big){}^2 \nonumber \\
&& +m_1\big(s_{12} \big(c_{34} s_{23}+e^{i \delta _{24}} c_{23} s_{24} s_{34}\big)+c_{12} \big(-e^{i \delta _{14}} c_{13} c_{24} s_{14} s_{34}-e^{i
\delta _{13}} s_{13} \big(c_{23} c_{34}-e^{i \delta _{24}} s_{23} s_{24} s_{34}\big)\big)\big){}^2 \nonumber \\
&& +e^{- i \alpha } m_2 \big(c_{12} \big(c_{34}
s_{23}+e^{i \delta _{24}} c_{23} s_{24} s_{34}\big)-s_{12} \big(-e^{i \delta _{14}} c_{13} c_{24} s_{14} s_{34}-e^{i \delta _{13}} s_{13} \big(c_{23}
c_{34} -e^{i \delta _{24}} s_{23} s_{24} s_{34}\big)\big)\big){}^2 \nonumber
\end{eqnarray}
\begin{eqnarray}
M_{es} &=& c_{14} \big(e^{i \delta _{14}} c_{24} c_{34} \big(e^{- i \gamma } m_4-e^{- i \alpha } c_{13}^2 m_2 s_{12}^2-e^{- i \beta } m_3 s_{13}^2\big)
s_{14}-e^{i \delta _{13}} c_{13} \big(e^{- i \beta } m_3-e^{- i \alpha } m_2 s_{12}^2\big) s_{13}  \nonumber \\
&& \big(e^{i \delta _{24}} c_{34} s_{23} s_{24}+c_{23}
s_{34}\big)+c_{12} c_{13} \big(m_1-e^{- i \alpha } m_2\big) s_{12} \big(e^{i \delta _{24}} c_{23} c_{34} s_{24}-s_{23} s_{34}\big) \nonumber \\
&& -c_{12}^2
c_{13} m_1 \big(e^{i \delta _{14}} c_{13} c_{24} c_{34} s_{14}-e^{i \delta _{13}} s_{13} \big(e^{i \delta _{24}} c_{34} s_{23} s_{24}+c_{23} s_{34}\big)\big)\big) \nonumber
\end{eqnarray}
\begin{eqnarray}
M_{\mu s} &=& e^{i \big(2 \gamma +2 \delta _{14}-\delta _{24}\big)} c_{14}^2 c_{24} c_{34} m_4 s_{24}+e^{i \big(2 \beta +\delta _{13}\big)} m_3 \big(e^{i
\delta _{13}} c_{13} c_{24} s_{23}-e^{i \big(\delta _{14}-\delta _{24}\big)} s_{13} s_{14} s_{24}\big) \nonumber \\
&& \big(-e^{-i \big(\delta _{13}-\delta
_{14}\big)} c_{24} c_{34} s_{13} s_{14}-c_{13} \big(e^{i \delta _{24}} c_{34} s_{23} s_{24}+c_{23} s_{34}\big)\big)+m_1 \big(-c_{23} c_{24}
s_{12}+c_{12} \big(-e^{i \delta _{13}} c_{24} s_{13} s_{23} \nonumber \\
&&  -e^{i \big(\delta _{14}-\delta _{24}\big)} c_{13} s_{14} s_{24}\big)\big)\big(s_{12}
\big(e^{i \delta _{24}} c_{23} c_{34} s_{24}-s_{23} s_{34}\big)+c_{12} \big(-e^{i \delta _{14}} c_{13} c_{24} c_{34} s_{14}+e^{i \delta _{13}}
s_{13} \big(e^{i \delta _{24}} c_{34} s_{23} s_{24} \nonumber \\
&& +c_{23} s_{34}\big)\big)\big)+e^{- i \alpha } m_2 \big(c_{12} c_{23} c_{24}+s_{12}\big(-e^{i
\delta _{13}} c_{24} s_{13} s_{23}-e^{i \big(\delta _{14}-\delta _{24}\big)} c_{13} s_{14} s_{24}\big)\big) \big(c_{12} \big(-e^{i \delta
_{24}} c_{23} c_{34} s_{24} \nonumber \\
&&+s_{23} s_{34}\big)+s_{12} \big(-e^{i \delta _{14}} c_{13} c_{24} c_{34} s_{14}+e^{i \delta _{13}} s_{13} \big(e^{i
\delta _{24}} c_{34} s_{23} s_{24}+c_{23} s_{34}\big)\big)\big) \nonumber
\end{eqnarray}
\begin{eqnarray}
M_{\tau s} &=& e^{ i \big(-\gamma +2\delta_{14}\big)} c_{14}^2 c_{24}^2 c_{34} m_4 s_{34}+e^{ i \big(-\beta +2 \delta_{13}\big)} m_3 \big(-e^{-i \big(\delta
_{13}-\delta _{14}\big)} c_{24} c_{34} s_{13} s_{14}-c_{13} \big(e^{i \delta _{24}} c_{34} s_{23} s_{24}+c_{23} s_{34}\big)\big)  \nonumber \\
&& \big(-e^{-i
\big(\delta _{13}-\delta _{14}\big)} c_{24} s_{13} s_{14} s_{34}+c_{13} \big(c_{23} c_{34}-e^{i \delta _{24}} s_{23} s_{24} s_{34}\big)\big)+m_1
\big(s_{12} \big(e^{i \delta _{24}} c_{23} c_{34} s_{24}-s_{23} s_{34}\big) \nonumber \\
&& +c_{12} \big(-e^{i \delta _{14}} c_{13} c_{24} c_{34} s_{14}+e^{i
\delta _{13}} s_{13} \big(e^{i \delta _{24}} c_{34} s_{23} s_{24}+c_{23} s_{34}\big)\big)\big) \big(s_{12} \big(c_{34} s_{23}+e^{i \delta
_{24}} c_{23} s_{24} s_{34}\big) \nonumber \\
&& +c_{12} \big(-e^{i \delta _{14}} c_{13} c_{24} s_{14} s_{34}-e^{i \delta _{13}} s_{13} \big(c_{23} c_{34}-e^{i
\delta _{24}} s_{23} s_{24} s_{34}\big)\big)\big)+e^{- i \alpha } m_2 \big(c_{12} \big(-e^{i \delta _{24}} c_{23} c_{34} s_{24}+s_{23} s_{34}\big) \nonumber \\
&& +s_{12}
\big(-e^{i \delta _{14}} c_{13} c_{24} c_{34} s_{14}+e^{i \delta _{13}} s_{13} \big(e^{i \delta _{24}} c_{34} s_{23} s_{24}+c_{23} s_{34}\big)\big)\big)
\big(-c_{12} \big(c_{34} s_{23}+e^{i \delta _{24}} c_{23} s_{24} s_{34}\big) \nonumber \\
&& +s_{12} \big(-e^{i \delta _{14}} c_{13} c_{24} s_{14} s_{34} -e^{i
\delta _{13}} s_{13} \big(c_{23} c_{34}-e^{i \delta _{24}} s_{23} s_{24} s_{34}\big)\big)\big) \nonumber
\end{eqnarray}
\begin{eqnarray}
M_{ss} &=& e^{- i \big(\gamma +\delta _{14}\big)} c_{14}^2 c_{24}^2 c_{34}^2 m_4+e^{ i \big(-\beta + 2 \delta_{13}\big)} m_3 \big(e^{-i \big(\delta
_{13}-\delta _{14}\big)} c_{24} c_{34} s_{13} s_{14}+c_{13} \big(e^{i \delta _{24}} c_{34} s_{23} s_{24}+c_{23} s_{34}\big)\big){}^2 \nonumber \\
&& +m_1 \big(s_{12}
\big(e^{i \delta _{24}} c_{23} c_{34} s_{24}-s_{23} s_{34}\big)+c_{12} \big(-e^{i \delta _{14}} c_{13} c_{24} c_{34} s_{14}+e^{i \delta _{13}}
s_{13} \big(e^{i \delta _{24}} c_{34} s_{23} s_{24} +c_{23} s_{34}\big)\big)\big){}^2 \nonumber \\
&&+e^{- i \alpha } m_2 \big(c_{12} \big(-e^{i \delta _{24}}
c_{23} c_{34} s_{24}+s_{23} s_{34}\big)  \nonumber \\
&& +s_{12} \big(-e^{i \delta _{14}} c_{13} c_{24} c_{34} s_{14}+e^{i \delta _{13}} s_{13} \big(e^{i \delta
_{24}} c_{34} s_{23} s_{24}+c_{23} s_{34}\big)\big)\big){}^2 \nonumber
\end{eqnarray}

\end{widetext}}

\end{document}